\documentclass[journal]{IEEEtran}
\usepackage{amsmath,amsfonts,amssymb}
\usepackage{bm}
\usepackage{algorithm}
\usepackage{algpseudocode}
\usepackage{array}
\usepackage[caption=false,font=footnotesize]{subfig}
\usepackage{textcomp}
\usepackage{stfloats}
\usepackage{tabularx}
\usepackage{url}
\usepackage{cite}
\usepackage{xcolor}
\usepackage[short]{optidef}
\usepackage[pdftex]{graphicx}
\usepackage{comment}
\usepackage{autobreak}
\usepackage[inline]{enumitem}
\usepackage{svg}
\usepackage{alphalph,etoolbox}
\newcommand{\T}{\mathcal{T}}
\newcommand{\C}{\mathcal{C}}
\newcommand{\SD}{\mathcal{S}}

\patchcmd{\subequations}{\alph{equation}}{\alphalph{\value{equation}}}{}{}

\usepackage{cellspace}
\allowdisplaybreaks[4]
\begin{document}

\title{Meeting Future Mobile Traffic Needs by Peak-Throughput Design of Next-Gen RAN}

\author{\IEEEauthorblockN{Paolo Fiore$^*$, Ilario Filippini$^*$ and Danilo De Donno$^\mathsection$}
\vspace{0.2cm}
\IEEEauthorblockA{
\hfill\\
$^*$ANTLab - Advanced Network Technologies Laboratory, Politecnico di Milano, Milan, Italy \\
$^\mathsection$Milan Research Center, Huawei Technologies Italia S.r.l, Milan, Italy\\
Email: $\{$paolo.fiore, ilario.filippini$\}$@polimi.it, danilo.dedonno@huawei.com
}
\vspace{-10mm}
\thanks{}
}

\markboth{Submitted to IEEE Transactions on Mobile Computing}%
{}


\maketitle

\begin{abstract}
{
\color{black}
Growing congestion in current mobile networks necessitates innovative solutions. This paper contributes a novel network planning approach for mmWave 5G networks in urban settings, focusing on Integrated Access and Backhaul (IAB) and the Smart Radio Environment (SRE). The mmWave traffic will be mainly made of short bursts to transfer large volumes of data and long idle periods where data are processed, necessitating changes in how mobile radio networks are designed. Our proposed optimization models integrate IAB with SRE technologies while leveraging the maximization of achievable peak throughput rather than conventional average throughput metrics. Results highlight the advantages of this approach during the network planning phase, providing insights into better accommodating the demands of mobile traffic without sacrificing the overall network capacity.
}

\end{abstract}


\section{Introduction}
\label{sec:introduction}
\IEEEPARstart{I}{n} the domain of mobile radio access networks (RAN), we are witnessing a growing spectrum congestion due to a surge of services and users. It is projected that by 2030 \cite{ericssonReportJun2025}, the overall mobile data usage will more than triple compared to 2023, and this growth will be primarily caused by 5G New Radio (NR) connections, overshadowing the utilization of previous mobile radio generations. To address this issue, every cellular network generation has been followed by another constantly aiming at improving spectral efficiency and bandwidth of the previous wireless communication standard. In this context, allocating new spectrum portions enables the use of larger bandwidths, which straightforwardly deliver a linear increase of achievable bitrates.

Following this logic, the 5th Generation (5G) standard for mobile communication of the 3rd Generation Partnership Project (3GPP) started to consider multiple frequency ranges. Frequency Range 1 (FR1), which partially overlaps with 4G frequencies; Frequency Range 2 (FR2) or mmWaves, operating at frequencies from 24 GHz up to 70 GHz, which permits bandwidth allocations of up to 2000 MHz; Frequency Range 3 or upper mid-band, operating at frequencies from 7 to 24 GHz, which is gaining more and more momentum in the transition towards the 6th Generation (6G), but not yet included in the standard; the Terahertz Frequency Range, with frequencies above 90 GHz, which promises astonishing throughputs for the 6G, but still requires to face nontrivial technical challenges.

The huge FR2 bandwidth promises a tenfold or more improvement in the bitrate. This drastic advancement paves the way for previously unattainable transmission speeds for end-users, unlocking possibilities for high-speed, low-latency applications envisioned in 5G. Simultaneously, this progress serves as a catalyst for innovation and advancement in industry and academic research environments. Consequently, mmWaves are recognized as a pivotal component in realizing the full potential of 5G and 5G-Advanced as outlined by 3GPP.  

The transition to mmWaves, however, comes with a set of challenges. Path loss increases with frequency, resulting in reduced network coverage. Furthermore, mmWaves interact poorly with common urban materials, experiencing remarkable penetration losses, negligible diffraction (e.g., from brick and concrete), and strong reflection (e.g., from glass and metal). As a result, FR2 communication predominantly relies on Line-of-Sight (LoS) paths. Yet, the complex urban layout can obstruct LoS between a Base Station (BS) and users, leading to the emergence of coverage blind spots. One viable approach to mitigate signal attenuation is \textit{network densification} which, by deploying more BSs, aims to increase the received power per user and the probability of LoS conditions. However, trenching and cabling costs may rapidly become a big issue. To address this, 3GPP Release 16 introduced a novel approach for cellular network coverage known as \textit{Integrated Access and Backhaul (IAB)}~\cite{3GPPTR38.874}. This new paradigm employs a wireless multi-hop backhaul where the signal is relayed from BS to User Equipment (UE) via IAB nodes, preserving the need of a single cabled connections in the cell.

The wireless multi-hop backhaul not only helps to mitigate path losses but also to alleviate the impact of static obstacles (like buildings) within the service area by strategically selecting paths that circumvent those obstacles. However, in urban settings, the challenges go beyond static obstacles: nomadic impediments, like vehicles and pedestrians, cause temporary obstructions on the wireless link ~\cite{8254900}. 

As a candidate to provide an effective solution to the impact of nomadic obstacles, the concept of the \textit{Smart Radio Environment (SRE)} has garnered substantial interest. A SRE implements the vision of a propagation environment no longer acting as an entity transmitter and receiver 
can only measure and adapt to, but rather a proper component of the system that can be actively tuned and utilized. The realization of an SRE implies integrating into the RAN one or many special radio devices, named \textit{Smart Radio Devices (SRD)}, capable of counteracting the channel impairments caused by nomadic obstacles. 

The two main types of SRD are: \textit{Reconfigurable Intelligent Surfaces (RISs)} and \textit{Network-Controlled Repeaters (NCRs)}. An RIS consists in planar array of passive elements such that, by applying a proper phase shift to each element, an impinging radio wave can be reflected to the desired direction, while minimizing power dispersion and interference. RIS's appeal lies in their quasi-passive nature, functioning as passive beamformers, and thus included as a subject of study within European Telecommunications Standards Institute (ETSI)~\cite{ETSIGRRIS001,ETSIGRRIS002,ETSIGRRIS003}. NCRs, recently introduced in 3GPP Release 18~\cite{3GPPTR38.867}, operate, instead, in an amplify-and-forward manner. Although entailing larger costs than an RIS, the power boost of NCR amplification offers more reliable links and faster throughputs.

While numerous studies have explored the impact of these technologies at the link and system levels, only a limited number have ventured into examining their broader implications across a urban RAN deployment~\cite{Moro2021,9771934,9930587,10293769}. In this article, we want to investigate whether the interoperation of mmWaves, IAB, RIS, and NCR can yield advantages within an urban RAN setup, and to what extent. By means of Mixed-Integer Linear Programming (MILP) network planning models we can identify and assess the optimal network layout and the best operational parameters in every potential deployment scenario. This allowed us to derive when and how this interplay can be effective, and thus provide reliable deployment guidelines to maximize the benefit/cost ratio of these devices.

\begin{figure*}[!ht]
    \centering
    \includegraphics[width=0.80\textwidth]{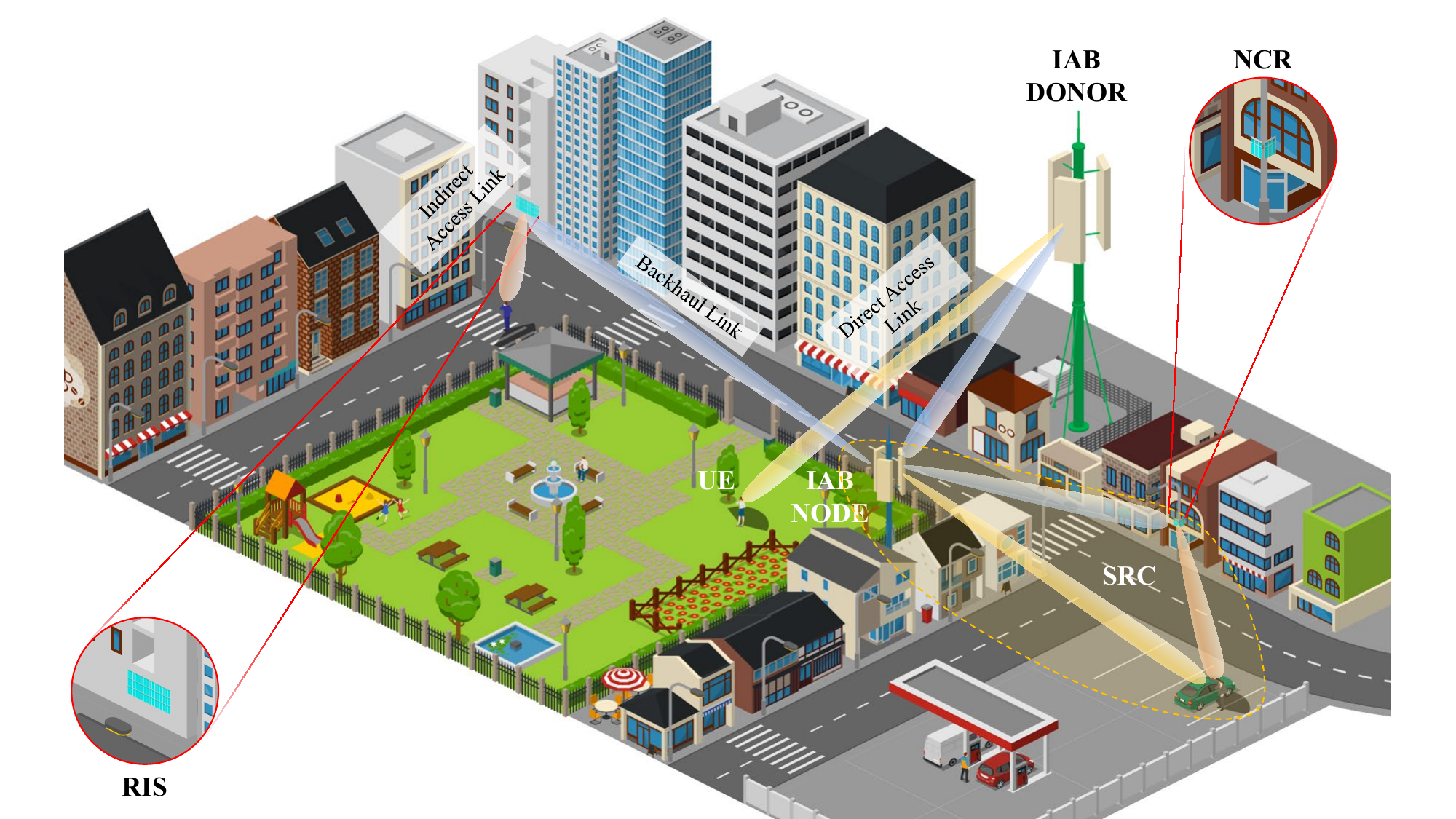}
    \caption{System model: the radio devices involved in the RAN and their relationships in a urban scenario.}
    \label{fig:Sysmod}
\end{figure*}

To better identify the effectiveness of a given network layout, we believe it is crucially important to understand how the network will be used in practice. Conventional network planning practices primarily focus on maximizing the \textit{average throughput per user}~\cite{6678102}. However, recent surveys~\cite{sandvineReport2024} indicate a growing dominance of video traffic, estimated to account for over 80\% of global mobile traffic. Video traffic exhibits peak, bursty characteristics~\cite{rao2011network}. At the same time, the very high throughput supported by mmWave communications allows the transfer of large volumes of bits in very short activation periods. That generates traffic patterns characterized by peak rate traffic injections followed by long idle periods to consume received data. Consequently, optimizing the achievable \textit{peak throughput} per user emerges as a superior approach when designing a network capable of accommodating mmWave traffic bursts. In addition, it brings the advantage of providing a network with the intrinsic capability of showing better speed-test performance, which is a key performance indicator to assess the quality of mobile radio networks from a user perspective.

This study investigates deployment guidelines for a 5G IAB-enabled mmWave RAN enriched with SRDs in an urban scenario featuring both static and nomadic obstacles, as the one in Figure~\ref{fig:Sysmod}. Results indicate that pursuing the optimization of peak throughput, as opposed to the conventional max-throughput approach, can be advantageous in better accommodating traffic bursts characterizing such networks.

{
\color{black}
The main contributions of this paper are as follows:
\begin{enumerate}
    \item We introduce a new network planning methodology for mmWave IAB networks that prioritizes \textit{peak throughput} rather than average throughput, effectively accommodating the bursty nature of modern mobile traffic.
    \item that seamlessly \textit{integrate IAB and SRE technologies} (RIS and NCR), enabling a comprehensive evaluation of their combined performance benefits.
    \item We conduct a detailed investigation into the effects of  
    static and nomadic obstacles in urban mmWave access links.
    \item We provide quantitative deployment guidelines to maximize network throughput performance and optimize the implementation of these emerging technologies in urban settings.
\end{enumerate}
}
 \textcolor{black}{In Table~\ref{tab:related_work_comparison}, we report a comparison of different planning methods and objectives found in the literature and this current work.}
The remainder of this paper is structured as follows: Section~\ref{sec:related_works} reviews the relevant literature related to our study, Section~\ref{sec:system_model} provides an in-depth survey of the various components within the considered network scenario and their interconnections. Section~\ref{sec:blockage_model} outlines the model we propose to account for the presence of static and nomadic obstacles. Section~\ref{sec:opt_models} delves into the MILP network planning models devised for this study. Section~\ref{sec:results} discusses the results of the numerical analysis we performed. The paper concludes with final remarks in Section~\ref{sec:conclusion}.

\section{Related Works}
\label{sec:related_works}
The potential and challenges of mmWaves for mobile radio networks have been well documented in the past decade: stimulated by the availability of large bandwidths and large antenna arrays in small form-factors, ~\cite{6515173,Akdeniz2014,rappaport2014,niu2015survey} have pioneered the study of mmWave propagation in indoor and urban environments and its channel modeling. ~\cite{8254900} highlights the challenges due to the presence of pedestrians in the area, while ~\cite{7342886} underlines the importance of beamforming techniques to make mmWaves suitable for cell coverage. ~\cite{8373698} provides a comprehensive survey of mmWave communications and guidelines to optimize physical, MAC, and network layers; ~\cite{7959169} addresses mmWave mobility issues and proposes mitigation solutions. Finally,
~\cite{8700132,sakaguchi2017and} discuss the integration of mmWave in the 5G NR context.

IAB in the context of mmWaves has garnered the interest of the scientific community since its inclusion in 3GPP as a study item. In~\cite{8514996,9040265}, the authors propose a full-stack ns-3 simulator of an IAB network operating at mmWaves; \cite{8882288} carry out a stochastic geometry analysis to show IAB deployments are capacity-limited by the overall traffic routed through the cabled connection. The authors in~\cite{9187867} investigate the performance of IAB in urban areas in terms of robustness and compare it to fiber-only and hybrid deployments, while~\cite{10.3389/frcmn.2021.647284} presents a survey on different applications of IAB with cache-enabled, optical transport, and non-terrestrial communication networks. In~\cite{9433519} discusses the advantages of IAB, identifying it as an easy-to-deploy solution for network densification.

RISs have recently emerged as a disruptive technology for wireless communications:~\cite{di2019smart},~\cite{8796365} and~\cite{tsilipakos2020toward} assess the revolutionary role RIS could hold in the SRE in its different modes of operation and use cases. Authors of~\cite{9086766} analyze the differences between reflectarrays and metasurfaces, and the challenges in their optimization and precoding, while in~\cite{9122596} attention is brought to the need for different phase-shift optimizations based on different network-level goals.

MmWave NCRs are another recent discussion topic: in~\cite{9119122}, the authors carry out a preliminary comparison between RIS and relays, identifying the RIS size as a determinant factor to approach the throughput performance of a relay; ~\cite{10065985} underlines the importance of side control information coming from the parent gNB to allow the NCR to be a better and more efficientalternative to legacy RF repeaters. In the system-level analysis carried out in~\cite{da2023system}, results show how the user throughput improves when NCRs are added, while intercell interference, though still present, does not cause outages;~\cite{9720231,10168854} highlights the combined use of both RIS and NCR as the best option for achieving coverage in urban scenarios.

\begin{table*}[!t]
\centering
\scriptsize
\color{black}
\caption{Comparison of Network Planning Approaches}
\label{tab:related_work_comparison}
\begin{tabularx}{\textwidth}{|c|>{\raggedright\arraybackslash}X|>{\raggedright\arraybackslash}X|>{\raggedright\arraybackslash}X|>{\raggedright\arraybackslash}X|>{\raggedright\arraybackslash}X|c|}
\hline
\textbf{Paper} & \textbf{Method/Approach} & \textbf{Key Technology Focus} & \textbf{Main Contribution} & \textbf{Optimization Objective} & \textbf{Network Scenario} & \textbf{Year} \\
\hline
Zhang et al.~\cite{zhang2021design} & Hybrid precoding with self-interference cancellation & Full-duplex mmWave IAB & Self-interference mitigation using zero-forcing and MMSE baseband combiner & Spectral efficiency maximization & Single-hop FD-IAB networks & 2021 \\
\hline
Albanese et al.~\cite{albanese2022ris} & Double-nested block coordinate ascent (RISA) & RIS-aware indoor planning & RIS placement optimization to solve dead-zone problems & Coverage optimization with interference management & Indoor railway station (Rennes) & 2022 \\
\hline
Zhang \& Filippini~\cite{zhang2024mobility} & Multi-Agent Reinforcement Learning (MARL) & mmWave IAB with mobility & Adaptive resource allocation under user/obstacle mobility & User throughput maximization via flow routing and scheduling & Multi-hop IAB with mobile users and obstacles & 2024 \\
\hline
Hervis Santana et al.~\cite{santana20245g} & Machine Learning + Genetic Algorithm & mmWave network planning & ML-based path loss approximation for fast network planning & Minimum AP count with coverage constraints & Indoor 28~GHz networks & 2024 \\
\hline Ayoubi et al.~\cite{ayoubi2025optimal} & Mixed-integer optimization (FCMC/MBCC) & Heterogeneous Smart Radio Environment & Joint planning of RIS, NCR, and STAR-RIS components & Cost minimization and coverage maximization & Urban high-frequency RAN & 2025 \\
\hline
\textbf{Our Work} & \textbf{MILP with peak-throughput formulation} & \textbf{mmWave IAB + Smart Radio Environment} & \textbf{Peak-throughput optimization for bursty traffic patterns} & \textbf{Peak user throughput maximization} & \textbf{Urban mmWave with static/nomadic obstacles} & \textbf{2025} \\
\hline
\end{tabularx}
\end{table*}

To the best of our knowledge, this is the first work to consider network planning in mmWave 5G FR2 settings with IAB and SRDs like RISs and NCRs. On top of that, the contributions of this article are:
\begin{itemize}
    \item A comprehensive and detailed MILP network planning optimization model, taking into account device installation, traffic routing, IAB multi-hop radio relaying, SRDs and capacity impairments due to obstacles;
    \item Different objective functions tailored to compare the standard, baseline approach of max-throughput maximization to the proposed peak-throughput optimization;
    \item A probabilistic blockage model integrating both self-blockage and nomadic obstacles in the cell to synthesize the probability of being in a specific state of blockage; 
    \item A 3D static blockage scenario derived from real data of the building distribution of the city of Milan, Italy.
\end{itemize}

\section{System Model}
\label{sec:system_model}
We are considering a 5G NR IAB network{\color{black} (\cite{3GPPTR38.874}\cite{8514996,9040265,8882288,9187867,10.3389/frcmn.2021.647284,9433519})} operating at mmWave frequencies (\cite{8254900}\cite{6515173,Akdeniz2014,rappaport2014,niu2015survey,7342886,8373698,7959169,8700132,sakaguchi2017and}\cite{dai2024user}), as shown in Figure~\ref{fig:Sysmod}, that includes:
\begin{enumerate*}[label={\textit{(\roman*)}}]
\item \textit{User Equipments (UE)}, i.e., hand-held devices such as smartphones or tablets,
\item a unique \textit{IAB Donor}, operating as a complete Base Station (BS) and the sole node in the RAN wired to the core network,
\item \textit{IAB Nodes}, which are simplified BS units capable of providing UEs with access and wirelessly relaying data to/from other IAB Nodes.
\end{enumerate*} 
The wireless backhaul links connecting IAB Nodes and the access links serving UEs both operate within the same frequency range (referred to as \textit{"in-band backhauling"}) at 28 GHz. We assume a tree-like IAB backhaul topology, as recommended in 3GPP specifications~\cite{3GPPTR38.874} to facilitate multi-hop end-to-end connections between the IAB Donor and UEs.

An asymmetric \textit{Time-Division Duplexing} (TDD) \textit{Downlink} (DL):\textit{Uplink} (UL) configuration 4:1 is employed, which is common to every 5G NR deployment. This configuration allows a maximum allocation of 80\% of frame resources for downlink data flows (\textit{from the network to the UEs}), and up to 20\% for the uplink ones (\textit{from the UEs to the network}).

\textit{Narrow beamforming} is available across all links to enhance signal propagation in the mmWave range. {\color{black} This, combined with the \textit{half-duplex} transmissions at each device, a significant path loss, and the consistent application of Time Division Multiplexing (TDM),
ensures that mutual interference is generally kept to a negligible level~\cite{7342886}. Consequently, the assumption is made that interference between different links is null, as in other works in the literature ~\cite{devoti2020, gupta2022system, 10649005, 7128403}}.

The incorporation of SRDs in an IAB Network is primarily driven by their ability to influence electromagnetic propagation to enhance channel conditions in the presence of obstacles. Operating as passive beamformers, RISs can effectively redirect incident radio signals toward specific directions. In contrast, NCRs are not passive; they possess the capability to amplify the received signal before retransmitting it to the intended receiver. However, at the system level, they play a role similar to RIS, albeit with a different implementation. 

Structurally, RISs {\color{black}(shown in the bottom-left corner of Figure~\ref{fig:Sysmod})} consist of planar surfaces of many small radiating elements {\color{black}\cite{ETSIGRRIS001,ETSIGRRIS002,ETSIGRRIS003} \cite{di2019smart,8796365,tsilipakos2020toward,9086766,9122596}}. The manipulation of phase shifts at each element enables the constructive alignment of incoming waves toward the intended direction and, at the same time, the capability to mitigate interference in other directions. Importantly, this process involves no power transmission; rather, it leverages the reflection of the received power. This is where the core aspect of \textit{passive beamformer} lies.
On the contrary, NCR {\color{black}~\cite{3GPPTR38.867}\cite{9119122,10065985,da2023system}}  {\color{black}(shown in the top-right corner of Figure~\ref{fig:Sysmod})} functions as an active amplify-and-forward relay, employing two back-to-back tunable phased-array panels capable of full-duplex operation\footnote{Full-duplex must be intended for the entire device as a whole: one panel can receive while the other transmits. The same panel cannot simultaneously transmit and receive.}. NCRs enhance the radio propagation environment by facilitating high-gain Non-Line-of-Sight (NLoS) paths between UEs and IAB Nodes. Given their amplification capability, NCRs generally offer higher bit rates compared to RISs. Similar to RISs, NCRs can be deployed to establish alternative paths for UEs, involving reflection and amplification.

Despite some differences, similar operational constraints related to their hardware limitations apply to both devices. RISs are defined by a specific field of view (FoV) within which both incoming and reflected radio wave directions must be confined. In the case of NCRs, when considering a single data flow direction (either downlink or uplink), one panel is consistently active as a transmitter, while the other concurrently operates as a receiver. Therefore, one panel (BS panel) is always perfectly oriented toward the parent BS, while the other one (UE panel) can serve UEs located in its FoV, similarly to an RIS. Full-duplex NCR transmissions are the results of recent technological advancements based on sophisticated interference cancellation and isolation techniques between the two panels. This introduces some limitations regarding the reciprocal orientation of the two NCR panels.

UEs can be served through either a direct link from an IAB Node (or the IAB Donor) or a \textit{Smart Radio Connection (SRC)}: a triplet involving a UE, an IAB Node, and an SRD (either an RIS or an NCR).
IAB Nodes are expected to dynamically control and modify SRC configurations~\cite{9720231}, activating the reflection through the SRD when obstacles obstruct the direct link. This improves the network's resilience to blockages caused by obstacles. The activation is based on SNR: when the selected link exhibits a lower SNR than its alternative, the serving BS dynamically switches its beam to the alternative link: this policy is named \textit{max-selection}. Note that, as in any radio planning approach, we do not delve into the details of link recovery operations. We assume, as discussed in the literature, that this mechanism is in place and effective\footnote{Note that efficient losses of this mechanism can be modeled by equivalent capacity losses in the access links.}. 

The mmWave channel model and path loss model, as presented in~\cite{10168854}, incorporates the effects of RIS and NCR, along with involved IAB Donor, IAB Nodes, and UEs. Reasonable considerations are made for obstacles in the deployment area based on the link type. Links connecting IAB Nodes (or IAB Donor) to other IAB nodes or SRDs, given the height of their sites, are only affected by static obstacles (e.g., buildings) and not by nomadic and random ground-level obstacles, like vehicles and pedestrians. Consequently, when LoS is available between those two types of devices, the associated bitrate is considered stable. Vice versa, links terminated in UEs, both direct (from an IAB Node) and reflected/relayed (from an SRD), are affected by static and nomadic obstacles. A detailed blockage model of these links, together with its impact on the average link bitrate, is provided in the next section.

\section{Blockage Model}
\label{sec:blockage_model}
A precise representation of the obstacles and their impact on link blockage becomes essential for planning and optimizing the deployment of mmWave RANs in order to ensure that the outputs of the analysis are consistent with real-life scenarios. To model the occurrence of link blockage, it is important to distinguish between three categories of blockages: 
\begin{enumerate*}[label={\textit{(\roman*)}}]
\item blockages due to static obstacles,
\item blockages due to nomadic obstacles, and 
\item self-blockage,
\end{enumerate*}
which we detail in the next paragraphs.

\subsection{Blockages due to static obstacles} \label{sub:static}
\begin{figure}[!ht]
\centering
\subfloat[Real-life building distribution]{\includegraphics[width=0.8\columnwidth]{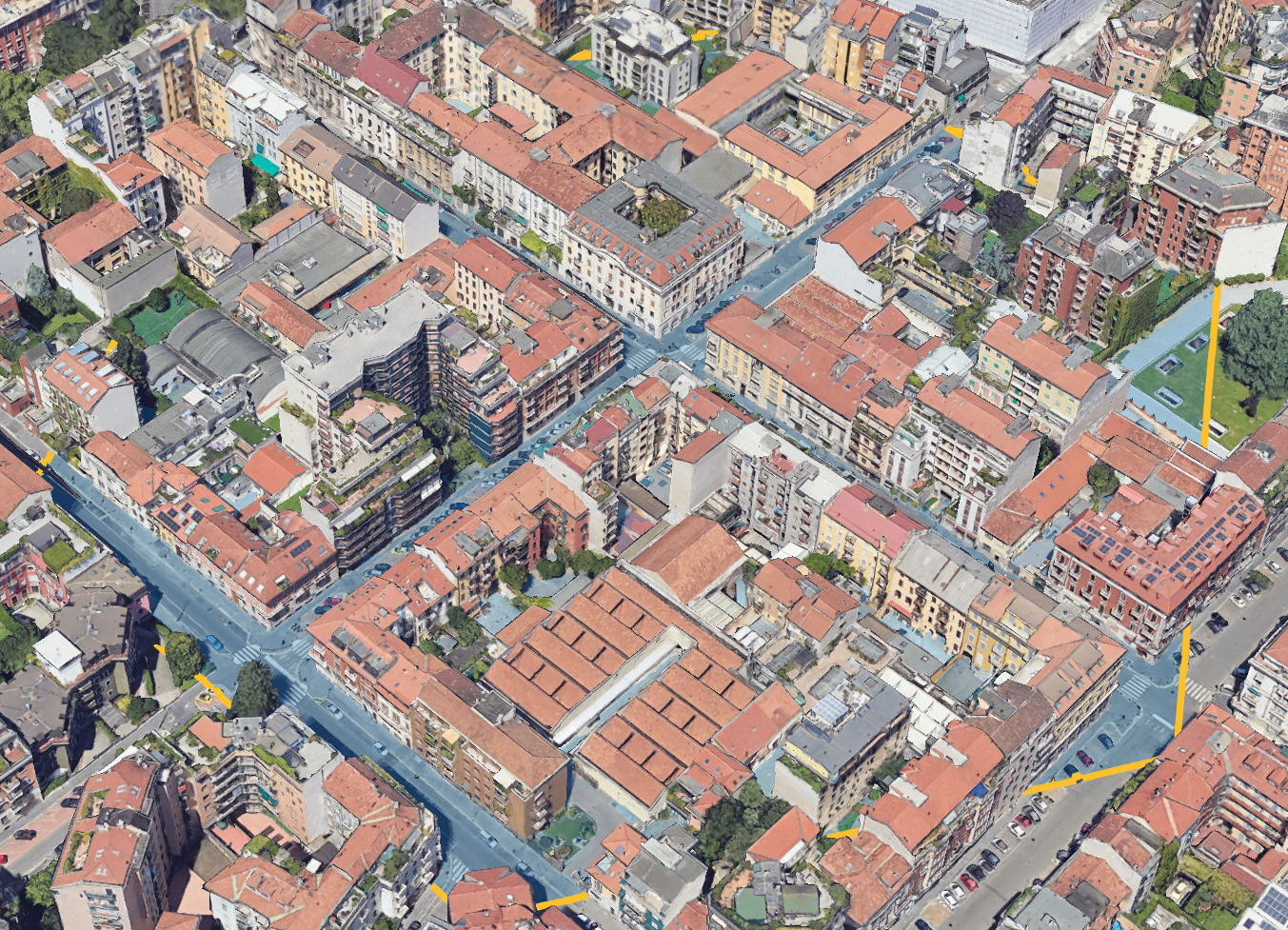}
\label{fig:real_buildings}}
\hfil
\subfloat[Polyhedral representation]{\includegraphics[width=0.8\columnwidth]{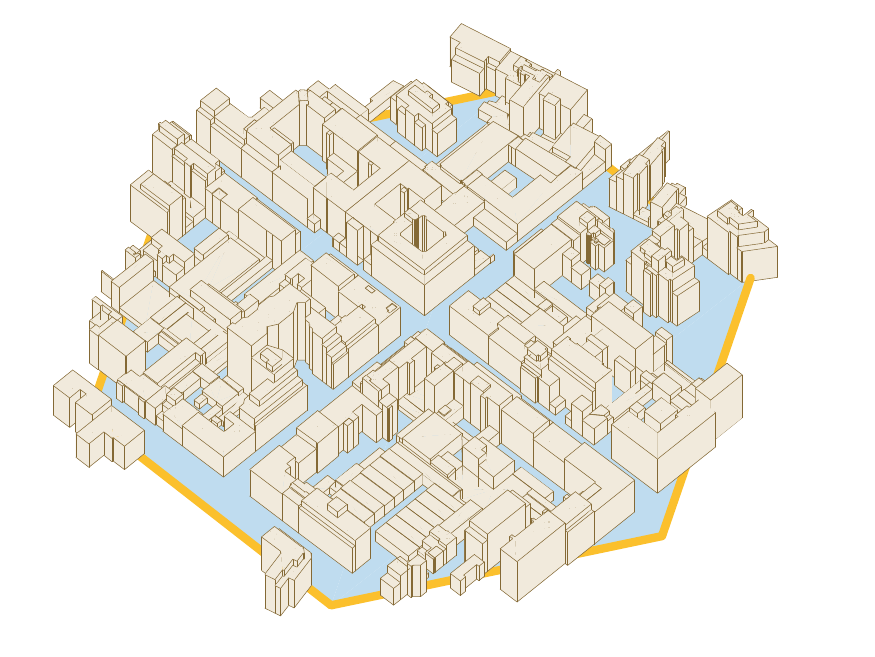}
\label{fig:polyhedra}}
\caption{Static blockage modeling. The blue-shaded area is where no static obstacle is present.}
\label{fig:building_processing}
\end{figure}

Static obstacles are typically represented by buildings or building structures. Their characterization is deterministic and relies on the geometry of the scenario under consideration, i.e., the positions of buildings and, on a smaller scale, street furniture; this information can be extracted from digital maps and pre-processed to identify areas where links are obstructed or unobstructed. An example of such a process is shown in Figure~\ref{fig:building_processing}: an infinite blockage loss is considered whenever a link intersects a building's polyhedron. This choice is justified by the high penetration loss of mmWave radio signals, exacerbated by the thickness of a building and the additional obstacles inside its volume. As a result, not only the identification of static blockage is deterministic, but also the link availability: all links crossing a static obstacle are considered unreliable and can be pruned in pre-processing, substantially reducing the solution space of the network planner.

\subsection{Blockages due to nomadic obstacles} \label{sub:nomadic}
Blockages due to nomadic obstacles are caused by moving vehicles or pedestrians{\color{black}~\cite{maccartney2016millimeter}} and are modeled using stochastic geometry methods~\cite{6290250,10168854}. When a nomadic obstacle blocks a link's LoS, the blockage loss can be computed with a modified version of the 3GPP blockage loss model~\cite{GRANW2018study}.

The blockage attenuation introduced by human blockage can reach as high as 20 dB if the mmWave link is subject to multiple human obstructions{\color{black}~\cite{collonge2004influence}}. However, the received signal quality may still be reliable enough to meet service requirements. For this reason, even if the signal is attenuated, the radio link can still be available for service and must be considered as part of a viable solution.

We rely on an improved version of the Blockage Model B in 3GPP TR 38.901 document~\cite{GRANW2018study}, which accurately characterizes the blockage loss affecting a direct communication link by means of geometric considerations. Specifically, each of the K potential obstacles, or blockers, obstructing the direct link between transmitter (Tx) and receiver (Rx) is modeled as a rectangular screen with height $h_k$ and width $w_k$. The blockage attenuation is computed according to a knife-edge diffraction model applied to the rectangle. Differently from 3GPP version, though, which only considers a two-dimensional geometry, our improved version  considers a {\color{black}2.5}-dimensional representation, which is more accurate when the blocker’s length cannot be ignored, f.i., in case of vehicles; {\color{black}such approach is validated by other works in the literature~\cite{jain2019impact, alwajeeh2017high,liu2013full}.}
This improved obstacle model is then implemented within a Monte-Carlo simulator in which random obstacles are dropped in a certain circle around the Tx: the centers of the blockers are distributed according to a homogeneous Poisson point process (PPP) on the 2D circle with radius $R$ in the $x-y$ plane with density $\lambda_B$, which denotes the mean number of blockers per square meter. The orientation of the blockers $\varrho_B$ is uniformly distributed at random, i.e., $\varrho_B \sim \mathcal{U}[0, 2\pi)$.

{\color{black}The empirical blockage probability $\hat{P}_B(R)=\frac{N_\text{B,R}}{N_\text{TOT,R}}$ as a function of the Tx-Rx distance $R$, is given by the number of blockage events $N_\text{B,R}$ over the total number of events considered $N_\text{TOT,R}$. $\hat{P}_B(R)$,} turns out to follow an empirical exponential law, and thus, an efficient and reasonable way to compute the blockage probability for any distance is to fit the exponential curve and interpolate it for the desired distance. The blockage probability due to nomadic obstacles, for specific obstacle size and density, can be written as:
\begin{equation}
    P_B(R)=1 - e^{(-\beta R)}
    \label{eq:fitted_exp}
\end{equation}
where $\beta$ is a positive real number. This behavior also complies with work in~\cite{6290250,6840343}.
The blockage loss cumulative distribution function CDF can be very accurately fitted by a \textit{Gaussian mixture model} of two Gaussian distributions, where the reciprocal weight depends on the height difference between blocker and BS.

\subsection{Self-blockage} \label{sub:self}
\begin{figure}
    \centering
    \includegraphics[width=0.80\columnwidth]{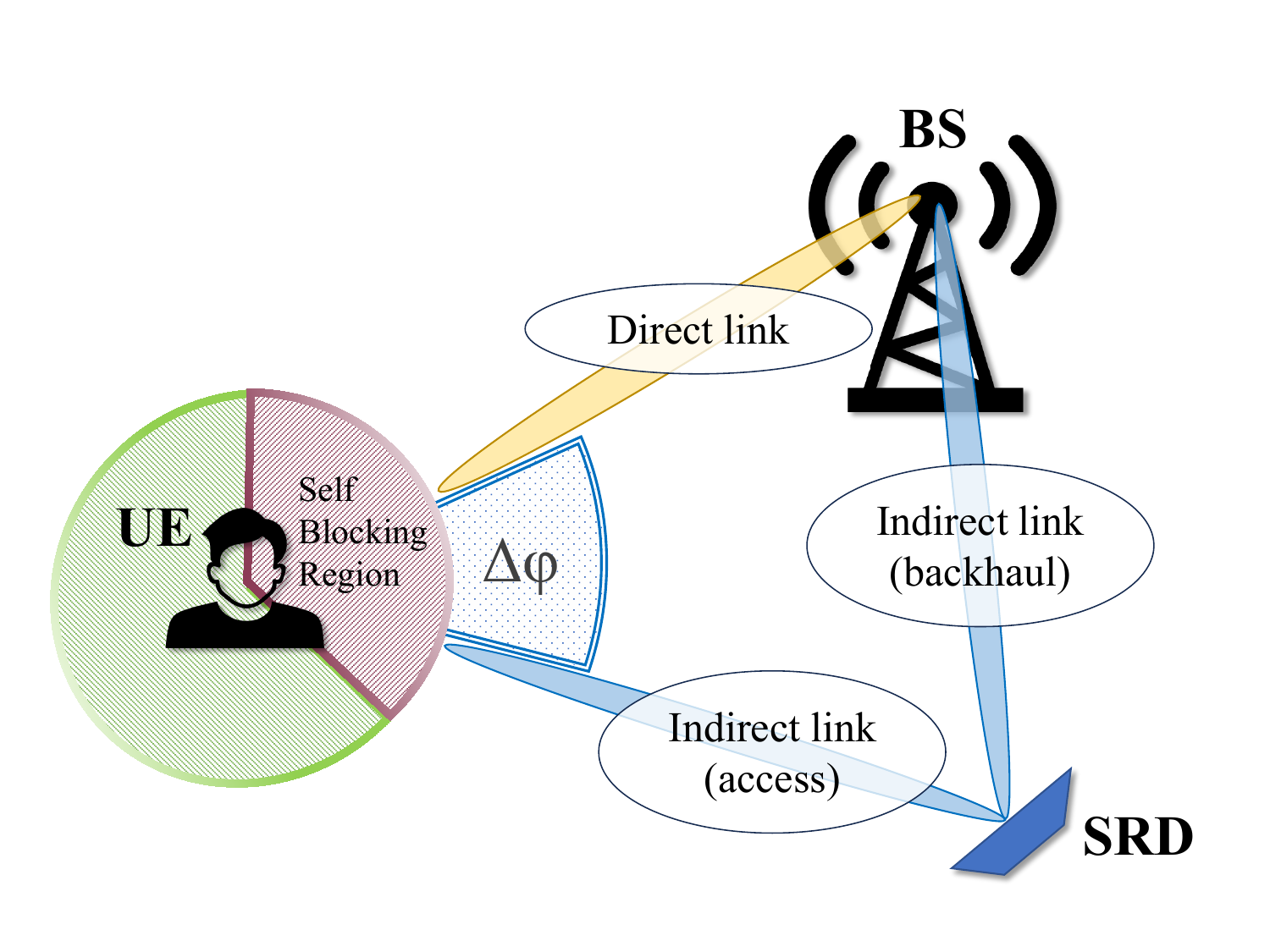}
    \caption{Schematic representation of the self-blocking region and how it could impair radio links.}
    \label{fig:SBR_scheme}
\end{figure}

Given its extreme proximity to the UE, the user body must be treated as a special obstacle that needs a dedicated approach to be correctly modeled. More precisely, the primary sources of blockage are the user's hand holding the device (e.g., a smartphone) and the user's body obstructing the LoS link to the serving BS~\cite{raghavan2019statistical}. Hence, self-blockage is contingent on both the UE's orientation and the orientation of the device itself.
Let's consider a UE being held by a person; if the person is at the center of a circle representing its surroundings, two circular sectors can be identified: one where the communication with another device is impaired by the person's body, known as the \textit{Self-Blockage Region (SBR)}, and one where it is not. In Fig.~\ref{fig:SBR_scheme} the interaction between SBR and radio access links is shown.
The way the device is held determines SBR size and orientation, as reported in Table 7.6.4.1-1 of 3GPP TR 38.901~\cite{GRANW2018study}. Table~\ref{tab:3GPP_table} lists the main SBR parameters according to the handheld device being used whether in portrait or landscape mode: $(x_{sb},y_{sb})$ are, respectively, the azimuth and the elevation angular width. For our study, we consider the reported SBR's azimuthal angle, $x_{sb}$, while, for the sake of simplicity, we conservatively set the SBR's elevation angle equal to $180^\circ$. Indeed, in a real scenario, when a link lies in the SBR's azimuthal range, it almost surely falls in the elevation range as well. In practice, a link can avoid SBR's elevation range only very close to the BS, but, at this location, connecting to the BS is difficult by itself due to the extreme tilt angles needed.

\begin{table}[H]
\centering

\caption{Self-Blockage Region parameters. \\
{\color{black}Ref.~Table 7.6.4.1-1 of 3GPP TR 38.901~\cite{GRANW2018study}.}}
\label{tab:3GPP_table}
 \begin{tabular}{|c|c|c|}
 \hline
  & $\bm{x_{sb}}$ & $y_{sb}$\\ [1ex] 
 \hline
 Portrait Mode & \textbf{120$^\circ$} & 80$^\circ$\\ [0.5ex] 
 \hline
 Landscape Mode & \textbf{160$^\circ$} & 75$^\circ$ \\ [0.5ex] 
 \hline
\end{tabular}
\end{table}

{\color{black} To properly assess the probability of a radio link falling within the SBR, we devise a set of equations leveraging basic aspects of statistical independence and conditional probabilities.} In a conventional single link access scenario, where the UE receives service through a direct link to the BS, the probability of this link lying within the SBR (denoted as event A) is given by:

\begin{equation}
P(A) = P_{p}\left(\frac{x_{sb}^p}{2 \pi}\right) + P_{l}\left(\frac{x_{sb}^l}{2 \pi}\right)
\label{eq:direct_sbr}
\end{equation}

Here, $P_p$ represents the probability of the user holding the UE in portrait mode, and $P_l = 1 - P_p$ is the probability of the user holding the UE in landscape mode. {\color{black} The fractions $\frac{x^\text{(.)}_\text{sb}}{2\pi}$ represent the angular sector falling within the SBR ($x^\text{(.)}_\text{sb}$) over the whole $2\pi$ circle.}

In a scenario where an SRD is present, a UE can be served by either the direct link or the link passing through the SRD, we refer to it as indirect link. In this case, the probability the indirect link lies within the SBR, given the direct link is in SBR as well, (Event B) depends on the angular distance between direct and indirect link . The probability of the event B is:

\begin{equation}
\begin{aligned}    
P(B) =~& P_{p}\left(\max\left(\frac{x_{sb}^p - \Delta\phi}{x_{sb}^p},0\right)\right)+\\
&~P_{l}\left(\max\left(\frac{x_{sb}^l - \Delta\phi}{x_{sb}^l},0\right)\right)
\label{eq:refl_sbr}
\end{aligned}
\end{equation}

where $\Delta\phi \in [0,\pi]$ represents the minimum angular distance between the direct and indirect link. When $\Delta\phi=0$, indicating overlapping links, the probability of the indirect link being blocked is $1$. As the angular spread $\Delta\phi$ increases, the indirect link leaves the SBR, and thus, the probability approaches zero. {\color{black} If $x_{sb}^{(.)} - \Delta\phi < 0$, the fraction of angular sector would be negative, therefore the probability is lower-bounded to zero.}

A similar rationale applies to the probability the indirect link lies outside the SBR,  given the direct link is not within the SBR (Event C), which can be expressed as:
\begin{equation}
\begin{aligned}
P(C) =~& P_{p}\left(\frac{2\pi - x_{sb}^p - \Delta\phi}{2\pi - x_{sb}^p}\right) +\\
~&P_{l}\left(\frac{2\pi - x_{sb}^l - \Delta\phi}{2\pi - x_{sb}^l}\right)
\label{eq:refl_nsbr}
\end{aligned}
\end{equation}

Regarding the attenuation due to self-blockage of a link within the SBR, observations in~\cite{raghavan2019statistical} indicate realistic attenuation values in the range from $5$ to $20$ dB, with a median of $15$ dB.

\subsection{Blockage probability due to non-static obstacles} \label{sub:states}
Leveraging the blockage probabilities outlined in Section~\ref{sub:nomadic} and Section~\ref{sub:self}, we formulate a comprehensive blockage probability model encompassing both nomadic blockage and self-blockage. Indeed, differently from static obstacles, nomadic obstacles and self-blockage require a probabilistic analysis of the blockage conditions. In addition, the simultaneous occurrence of both blockage types has a stronger impact than the case a single blockage arises, as the attenuation of the two obstacles is additive. As a consequence, any direct and indirect links of an SRC can be characterized by one of the following $4$ states of blockage:
\begin{itemize}
\item $\mathbf{D_{F}}$($\mathbf{I_{F}}$): the direct(indirect) link is free from obstacles,
\item $\mathbf{D_{N}}$($\mathbf{I_{N}}$): the direct(indirect) link is blocked by a nomadic obstacle,
\item $\mathbf{D_{SB}}$($\mathbf{I_{SB}}$): the direct(indirect) link is blocked by the self-blocking phenomenon, or
\item $\mathbf{D_{N,SB}}$($\mathbf{I_{N,SB}}$): the direct(indirect) link is simultaneously blocked by both nomadic obstacles and self-blocking.
\end{itemize}
Therefore, SRC can exist in one of $16$ different states, and the probability of each state, over a sufficiently long observation period, corresponds to the percentage of time the SRC experiences that specific state.

\begin{table}[t]
\centering
\caption{SBR-related events}
\label{tab:blockage_events}
   \begin{tabularx}{\columnwidth}{|c|c|X|}
 \hline
 \textbf{Event} & \textbf{Link status} & \textbf{Description} \\ [1ex] 
 \hline
 A & $\mathbf{D_{SB}}$ & Direct link is in the SBR, Eq.~\eqref{eq:direct_sbr}\\ [0.5ex] 
 \hline
 B & $\mathbf{(I_{SB} | A)}$ & Indirect link is in the SBR given that the direct link is  in the SBR, Eq.~\eqref{eq:refl_sbr}\\ [0.5ex] 
 \hline
 C & $\mathbf{(\neg I_{SB}| \neg A)}$ & Indirect link is not in the SBR, given that the direct link is not in the SBR, Eq.~\eqref{eq:refl_nsbr}\\ [0.5ex] 
 \hline
 D & $\mathbf{D_{N}}$ & Nomadic obstacle is present on the direct link, Eq.~\eqref{eq:fitted_exp}\\ [0.5ex] 
 \hline
 E & $\mathbf{I_{N}}$ & Nomadic obstacle is present on the indirect link, Eq.~\eqref{eq:fitted_exp}\\ [0.5ex] 
 \hline
\end{tabularx}
\end{table} 
We consider the blockage events in Table~\ref{tab:blockage_events}. Events $A$, $B$, and $C$ have been defined in the previous sections and their probabilities are given by (\ref{eq:direct_sbr}-\ref{eq:refl_nsbr}). Events $D$ and $E$ are straightforward events and their probabilities are computed by fitting the exponential distribution from a Monte-Carlo simulation, as indicated in~\eqref{eq:fitted_exp}. Note that we consider Event $D$ and Event $E$ mutually independent, and both independent of Events $A$, $B$, and $C$. By means of the probabilities of these blockage events, we compute the probability of each of the $16$ states an SRC can operate. Table~\ref{tab:16_states} illustrates these states, where the columns represent the four states of blockage for the direct link, and the rows represent those for the indirect link. The probabilities in each cell, when multiplied, yield the final probability of being in that particular SRC state\footnote{Note that, in case no SRD is present and only the direct link exists, the states collapse to the first row of the table. In this case, the indirect link is always free (as it doesn't exist): the probabilities it lies in an SBR (Event $B$) or is obstructed by a nomadic obstacle (Event $E$) are 0, and, consequently, the probability of not occurrence of the self-blockage on the indirect link (Event $C$) is equal to 1.}.

\textcolor{black}{These probabilities are then employed in Sec.~\ref{sub:cap} (see Eq.~\ref{eq:src_cap}) to compute the capacities of the potential SRCs, which are in turn a fundamental parameter of the network planning models subsequently defined (\ref{sub:mtf}, \ref{sub:ptf}).}

\begin{table*}[ht]
\centering
\caption{The 16 potential blockage states for the dual access links of a SRC}
\label{tab:16_states}
 \begin{tabular}{| c | l | l | l | l |} 
 \hline
  & \multicolumn{1}{c|}{$\mathbf{D_{F}}$} & \multicolumn{1}{c|}{$\mathbf{D_O}$} & \multicolumn{1}{c|}{$\mathbf{D_{SB}}$} & \multicolumn{1}{c|}{$\mathbf{D_{O,SB}}$} \\ [0.5ex] 
 \hline
 $\mathbf{I_{F}}$ & \textbf{1}:~$P(C) \neg P(A)  \neg P(E)   \neg P(D)$ & \textbf{2}:~$P(C) \neg P(A)  \neg P(E)   P(D)$ &\textbf{3}:~$\neg P(B)  P(A)  \neg P(E)   \neg P(D)$ & \textbf{4}:~$\neg P(B)  P(A)  \neg P(E)   P(D)$ \\ [0.1ex] 
 \hline
 $\mathbf{I_O}$ & \textbf{5}:$P(C) \neg P(A)  P(E)   \neg P(D)$ & \textbf{6}:$P(C) \neg P(A)   P(E)   P(D)$ & \textbf{7}:$\neg P(B)  P(A)   P(E)   \neg P(D)$ & \textbf{8}:$\neg P(B)  P(A)  P(E)   P(D)$ \\ [ 0.1ex] 
 \hline
 $\mathbf{I_{SB}}$ & \textbf{9}:$\neg P(C) \neg P(A)  \neg P(E)   \neg P(D)$ & \textbf{10}:$\neg P(C) \neg P(A)  \neg P(E)   P(D)$ & \textbf{11}:$P(B)  P(A)  \neg P(E)   \neg P(D)$ & \textbf{12}:$P(B)  P(A)  \neg P(E)   P(D)$ \\ [ 0.1ex]  
 \hline
 $\mathbf{I_{O,SB}}$ & \textbf{13}:$\neg P(C) \neg P(A)  P(E)   \neg P(D)$ & \textbf{14}:$\neg P(C) \neg P(A)   P(E)   P(D)$ & \textbf{15}:$P(B)  P(A)  P(E)   \neg P(D)$ & \textbf{16}:$P(B)  P(A)  P(E)   P(D)$ \\ [ 0.1ex] 
 \hline
\end{tabular}
\end{table*}

\section{Optimization Models}
\label{sec:opt_models}
We believe that a thorough investigation of the optimal network planning solutions in several different scenarios can provide a deep understanding of the phenomena impacting on optimal choices, which enables the design of effective network deployment guidelines. For this reason, accurately modeling the behavior and the limitations of all network players is pivotal to the entire study.

In this work we deal with two distinct MILP models that rely on the same network setup, but focus on different planning objectives: one is the \textbf{Mean-Throughput Formulation (MTF)}, a baseline formulation where the objective function is the maximization of the overall \textit{mean} user throughput, the other is the \textbf{Peak-Throughput Formulation (PTF)}, an extension of the baseline approach where the maximization addresses the \textit{peak} user throughput. 

Both formulations share a common notation, defined as follows. We define a set $\C$ of Candidate Sites (CS), where a network device (the IAB Donor, an IAB Node, an RIS, or an NCR) can be deployed to cover a planned mmWave network in an urban area. Within $\C$, two CSs, $\hat{c}$ and $\Tilde{c}$, are designated for the installation of two special devices: $\hat{c}$ is reserved for the IAB Donor, while $\Tilde{c}$ serves as a CS for a "fake" SRD, representing a null SRD. This approach lets the solver to determine whether to assign to each UE a real SRC (a BS, a UE, and an SRD) or a "fake" SRC (a BS, a UE, and the "fake" SRD), thus the availability of the sole direct link. This allows to reduce the formulation complexity as avoids introducing additional variables to characterize the two types of access connections.
Test Points (TP), which represent centroids of traffic mirroring the geographical distribution of the UEs in the area, are denoted by set $\T$. Moreover, set $\SD$ represents available SRD types. We consider two of them, namely, RIS and NCR. In case other device types or subtypes were required, this set can be easily extended to represent any type of device. 
\begin{table}[htbp]
\caption{Sets used in the optimization models}
\label{tab:sets}
\centering
 \begin{tabular}{|c|l|}
 \hline
 \textbf{Set} & \textbf{Description} \\ [1ex] 
 \hline
 $\T$ & Set of Test Points (TPs) \\ [0.5ex] 
 \hline
 $\C$ & Set of Candidate Sites (CSs) \\ [0.5ex] 
 \hline
 $\SD$ & Set of types of Smart Radio Devices (SRDs) \\ [0.5ex] 
 \hline
 $\{\hat{c}\}$ & CS reserved for the IAB Donor\\ [0.5ex] 
 \hline
 $\{\Tilde{c}\}$ & CS reserved for the Fake SRD\\ [0.5ex] 
 \hline
\end{tabular}
\end{table}
The above-mentioned sets are summarized in Table~\ref{tab:sets}.

All physical properties of an SRC, like SNR, blockage loss, and FoV, are captured by feasibility parameters $\Delta$ and capacity parameters $C \in \mathbb{R^+}$. Specifically, the parameter $\Delta_{t,c,r,s}^{\text{SRC}}$ is set to 1 when a potential SRC may exist between TP $t \in \T$, an IAB device (IAB Node or IAB Donor) in $c \in \C$, and an SRD of type $s$ in CS $r \in \C$. Likewise, the parameter $\Delta_{c,d}^{\text{BH}}$ denotes the availability of a potential in-band backhaul link between IAB nodes $c$ and $d$ in $\C$; in this case, though, only SNR and blockage loss considerations are taken into account, since the tri-sectorial nature of IAB devices provides them with a $360^\circ$ FoV. As for capacity considerations, $C_{t,c,r,s}^{\text{DL,SRC}}$ represents the average downlink capacity of the SRC defined by  TP $t $,  IAB device in $c$, and SRD of type $s$ in CS $r$. Correspondingly, $C_{t,c,r,s}^{\text{UL,SRC}}$ refers to the uplink capacity. The rationale followed to compute these capacities is detailed in subsection~\ref{sub:cap}. The parameter $C_{c,d}^{\text{BH}} \in \mathbb{R^+}$ characterizes the capacity of a backhaul link between nodes $c$ and $d$ in $\C$. A minimum demand of $D^{\text{DL}}$ ($D^{\text{UL}}$) must be met for each TP in downlink (uplink). Parameter $\alpha \in (0,1)$ indicates the fraction of wireless resources dedicated to downlink traffic in each radio device. It aims to capture 4:1 or 2:3 TDD frame structures typically used in 5G NR. Parameter $M$ is an upper bound for routed traffic in the planned RAN and it is used by constraint linearization techniques.

The set of network devices for deployment is limited by the budget, denoted as $B$. The prices for IAB Nodes, RISs, and NCRs are represented by $P^{\text{IAB}}$, $P^{\text{RIS}}$, and $P^{\text{NCR}}$, respectively\footnote{Since one and only IAB Donor is needed, we excluded it from the optimization deployment budget.}.

Each RIS is associated with a FoV angle $F$ where both the IAB device and the TP of the SRC must lie to use the RIS. For the NCR, $F$ refers to the FoV of the UE panel. To enforce FoV visibility constraints, parameters $\Phi_{r,t}^{\text{A}},\Phi_{r,c}^{\text{B}} \in \left[0,2\pi\right]$ must be defined. The former indicates the angle between SRD $r \in \C$ and TP $t \in \T$, while the latter represents the angle between SRD $r$ and BS $c \in \C$.
\begin{table}[htbp]
\caption{Parameters used in optimization models}
 \label{tab:pars}
\centering
 \begin{tabularx}{\columnwidth}{|Sc|Sc|X|} 
 \hline
 \textbf{Parameter} & \textbf{Domain} & \textbf{Description} \\ [1ex] 
 \hline
 $\Delta_{t,c,r,s}^{\text{SRC}}$ & $\{0;1\}$ & Coverage in SRC with SRD of type $s \in \SD$ involving CSs $c,r \in \C$ and TP $t \in \T$\\ [0.5ex] 
 \hline
 $\Delta_{c,d}^{\text{BH}}$ & $\{0;1\}$ & Coverage between CSs $c,d \in \C$ \\ [0.5ex] 
 \hline
  $C_{t,c,r,s}^{\text{DL,SRC}}$ & $\mathbb{R^+}$ & Average downlink capacity of access link from TP $t \in \T$, with BS $c$ and SRD of type $s \in \SD$ in CS $r$: $c,r \in \C$\\ [0.5ex] 
\hline
 $C_{t,c,r,s}^{\text{UL,SRC}}$ & $\mathbb{R^+}$ & Average uplink capacity of access link from TP $t \in \T$, with BS $c$ and SRD of type $s \in \SD$ in CS $r$: $c,r \in \C$\\ [0.5ex] 
\hline
 $C_{c,d}^{\text{BH}}$ & $\mathbb{R^+}$ & Capacity of backhaul link $(c,d)$, $c,d \in \C$ \\ [0.5ex]
\hline
 $D^{\text{DL}}$ & $\mathbb{R^+}$ & Downlink traffic demand for each TP\\ [0.5ex]
 \hline
  $D^{\text{UL}}$ & $\mathbb{R^+}$ & Uplink traffic demand for each TP\\ [0.5ex]
 \hline
 $\alpha$ & $\left[0,1\right]$ & percentage of time dedicated to downlink for each base station\\[0.5ex] 
\hline
 $B$ & $\mathbb{R^+}$ & Budget\\ [0.5ex]
 \hline
$P^{\text{IAB}}$ & $\mathbb{R^+}$ & Price for the installation of a IAB node in any CS\\ [0.5ex] 
 \hline
 $P^{\text{RIS}}$ & $\mathbb{R^+}$ & Price for the installation of an RIS in any CS\\ [0.5ex] 
 \hline
  $P^{\text{NCR}}$ & $\mathbb{R^+}$ & Price for the installation of an NCR in any CS\\ [0.5ex] 
 \hline
$\Phi_{r,t}^{\text{A}}$ & $\left[0,2\pi\right]$ &  angle between TP $t \in \T$ and CS $r \in \C$\\ [0.5ex]
\hline
$\Phi_{r,c}^{\text{B}}$ & $\left[0,2\pi\right]$ &  angle between CS $c,r \in \C$\\ [0.5ex]
\hline
$F$ & $\left[0,\pi\right]$ & maximum angle of the reflected/impinging radio wave on a SRD\\
\hline
$M$ & $\mathbb{R^+}$ & Big-M parameter representing the upper limit of the traffic routed through the Donor to the IAB network\\[0.5ex]
\hline
\end{tabularx}
\end{table}
All parameters used in the models are summarized in Table~\ref{tab:pars}.

{\begin{table}[htbp]
 \caption{Decision Variables used in optimization models}
 \label{tab:vars}
\centering
 \begin{tabularx}{\columnwidth}{|Sc|Sc|X|}
 \hline
 \textbf{Variable} & \textbf{Domain} & \textbf{Description} \\ [1ex] 
 \hline
 \multicolumn{3}{|c|}{\textbf{MTF variables}}\\
 \hline
 $y_c^{\text{DON}}$ & $\{0;1\}$ & IAB Donor installation of CS $c \in \C$\\ [0.5ex] 
 \hline
 $y_c^{\text{IAB}}$ & $\{0;1\}$ & IAB Node installation of CS $c \in \C$\\ [0.5ex] 
 \hline
 $y_c^{\text{RIS}}$ & $\{0;1\}$ & RIS installation of CS $c \in \C$\\ [0.5ex] 
 \hline
 $y_c^{\text{NCR}}$ & $\{0;1\}$ & NCR installation of CS $c \in \C$\\ [0.5ex] 
 \hline
 $x_{t,c,r,s}$ & $\{0;1\}$ & association of TP $t \in \T$ to BS $c \in \C$ and SRD $s \in \SD$ in CS $r \in \C$\\ [0.5ex] 
 \hline
 $z_{c,d}$ & $\{0;1\}$ & backhaul link activation between CSs $c,d \in \C$\\ [0.5ex] 
 \hline
 $f_{c,d}^{\text{DL}}$ & $\mathbb{R^+}$ & downlink backhaul traffic on link (c,d):  $ c,d \in \C$\\ [0.5ex] 
 \hline
  $f_{c,d}^{\text{UL}}$ & $\mathbb{R^+}$ & uplink backhaul traffic on link (c,d):  $ c,d \in \C$\\ [0.5ex] 
 \hline
$t_c^{\text{DL}}$ & $\left[0,1\right]$ & Fraction of time per resource unit dedicated to downlink traffic in CS $c \in \C$\\ [0.5ex] 
\hline
$t_c^{\text{UL}}$ & $\left[0,1\right]$ & Fraction of time per resource unit dedicated to uplink traffic in CS $c \in \C$\\ [0.5ex] 
\hline
 $\phi_c$ & $\left[0,2\pi\right]$ &SRC orientation of CS $c \in \C$\\ [0.5ex] 
 \hline
 $w_c^{\text{DL}}$ & $\mathbb{R^+}$ & Total traffic incoming in the IAB network through the Donor $c\in\C$ \\[0.5ex] 
 \hline
 $w_c^{\text{UL}}$ & $\mathbb{R^+}$ & Total traffic going out of the IAB network through the Donor $c\in\C$ \\[0.5ex] 
 \hline
 $g_{t,c,r,s}^{\text{DL}}$ & $\mathbb{R^+}$ & downlink demand guaranteed to SRC $(t,c,r)$ with device $s: t\in\T,c,r\in\C,s\in\SD$\\[0.5ex] 
 \hline
 $g_{t,c,r,s}^{\text{UL}}$ & $\mathbb{R^+}$ & uplink demand guaranteed to SRC $(t,c,r)$  with device $s: t\in\T,c,r\in\C,s\in\SD$\\[0.5ex] 
 \hline
 \multicolumn{3}{|c|}{\textbf{PTF additional variables}}\\
 \hline
 $f_{t,c,d}^{\text{X,DL}}$ & $\mathbb{R^+}$ & extra downlink traffic on link (c,d) for TP $t: c,d \in \C, t \in \T$\\ [0.5ex] 
 \hline
 $f_{t,c,d}^{\text{X,UL}}$ & $\mathbb{R^+}$ & extra uplink traffic on link (c,d) for TP $t: c,d \in \C, t \in \T$\\ [0.5ex] 
 \hline
 $w_{t,c}^{\text{X,DL}}$ & $\mathbb{R^+}$ & Extra downlink traffic to the IAB network through the Donor $c$ for TP $t: t\in\T,c\in\C$\\[0.5ex] 
 \hline
 $w_{t,c}^{\text{X,UL}}$ & $\mathbb{R^+}$ & Extra uplink traffic from the IAB network through the Donor $c$ for TP $t: t\in\T,c\in\C$\\[0.5ex] 
 \hline
 $g_{t,c,r,s}^{\text{X,DL}}$ & $\mathbb{R^+}$ & Extra downlink demand given to SRC $(t,c,r)$ with SRD $s\in\SD: t\in\T,c,r\in\C$ \\[0.5ex] 
 \hline
 $g_{t,c,r,s}^{\text{X,UL}}$ & $\mathbb{R^+}$ & Extra uplink demand given to SRC $(t,c,r)$ with SRD $s\in\SD: t\in\T,c,r\in\C$ \\[0.5ex] 
 \hline
\end{tabularx}
\end{table}} 
{\color{black} The final group of formulation terms includes \textbf{decision variables}, detailed in Table~\ref{tab:vars}. The values taken by those variables after the optimization describe the optimal solution of the network planning problem.} These variables indicate the installation and placement of devices ($y_c^{\text{DON}}, y_c^{\text{IAB}}, y_c^{\text{RIS}}, y_c^{\text{NCR}}$), the establishment of network links between them ($x_{t,c,r,s}, z_{c,d}$), the traffic flow through each link ($f_{c,d}^{\text{UL}}, f_{c,d}^{\text{DL}}, g_{t,c,r,s}^{\text{UL}}, g_{t,c,r,s}^{\text{DL}}, w_c^{\text{UL}}, w_c^{\text{DL}}$), the wireless resources dedicated by each IAB devices to UL or DL traffic ($t_c^{\text{DL}}, t_c^{\text{UL}}$), and the orientation of installed SRDs ($\phi_c$). More information about the meaning of these variables will be given in the next sections, where constraints are described.

\subsection{Computing SRC capacity}\label{sub:cap}
The generic capacity $C$ of a 5G NR link is given by the formula:

{\small
\begin{equation}
\label{eq:3gpp_cap_huawei}
C = (1-OH_{\text{5G}})\frac{(N_{\mu,B}^{\text{RB}}N_{sc}^{\text{RB}}N_{symb}^{\text{slot}}2^\mu N_{sf}^{\text{F}}Q_m\frac{R}{1024}\nu)}{t_F}    
\end{equation}}

Here, $N_{\mu,B}^{\text{RB}}$ is the number of resource blocks associated with a chosen numerology $\mu$ and bandwidth $B$ (refer to Table~\ref{tab:RE} in the Appendix), $N_{sc}^{\text{RB}}=12$ is the number of subcarriers per resource block, $N_{symb}^{\text{slot}}=14$ is the number of OFDM symbols per slot, $2^\mu$ is the number of slots per subframe, $N_{sf}^{\text{F}}=10$ is the number of subframes per frame, $t_F=10ms$ is the frame duration, $\nu$ is the number of MIMO layers, $OH_{\text{5G}}$ is the estimated 5G overhead, $Q_m$ is the modulation order (i.e. the number of bits per symbol), and $R$ is the code rate (i.e. the number of information bits over a total of 1024 bits). The values of $Q_m$ and $R$ are provided in Table~\ref{tab:MCS} in the Appendix. The most suitable $(Q_m,R)$-pair  is selected by comparing the SNR of the considered radio link with the SNR thresholds in the first column of Table~\ref{tab:MCS}.
In order to obtain the average link capacity for an SRC, we must average the achievable rate of each of the $16$ states described in Table~\ref{tab:16_states}. Eq.~\ref{eq:3gpp_cap_huawei} can be used to compute the link capacities of the direct link $C_{t,c,\Tilde{c},i}^{\text{DIR}}$ and indirect link $,C_{t,c,r,s,i}^{\text{IND}}$ of the SRC defined by BS in $c\in\C$, TP $t\in\T$ and SRD installed in $r\in\C$, when the SRC is in state $i$. Note that in the case of states with blockages, the proper blockage attenuation must be inserted in the link budget to compute the correct SNR. The final, long-term SRC capacity is then computed using the following formula:
{
\small
\begin{flalign}
    \label{eq:src_cap}
    &C_{\substack{t,c,\\r,s}}^{\text{SRC}} = \sum_{i=1}^{16}P_{\substack{t,c, \\ r,s,i}}\max(C_{\substack{t,c,\\ \Tilde{c},i}}^{\text{DIR}},C_{\substack{t,c,\\r,s,i}}^{\text{IND}}),\hspace{-1em}& \forall t\in\T,c,r\in\C,s\in\SD.
\end{flalign}
}
where $P_{t,c,r,s,i}$ is the probability of an SRC ($t,c,r$) with SRD of type $s\in\SD$ to be in blockage state $i\in\left[1,16\right]$ (Table~\ref{tab:16_states}). The formula implements the \textbf{max-selection} policy mentioned in Section~\ref{sec:system_model}: in each of the blockage states an SRC can operate, the link with the highest capacity between direct and indirect links is always selected, and its capacity is weighted by the occurrence probability of the blockage state $P_{(\cdot,i)}$. Note that in the case the access link between an IAB device and a UE does not make use of an SRD, the max function provides just the capacity of the direct link, given the indirect link doesn't exist and its capacity is $0$ Mb/s. The resulting parameter, $C_{t,c,r,s}^{\text{SRC}}$ is the average link capacity weighted by the probability of being in a specific blockage state. This process is repeated for both downlink and uplink.

\subsection{Mean-Throughput Formulation (MTF)}\label{sub:mtf}
We describe the proposed MTF in the following, starting from its objective function and proceeding with its constraints grouped according to their purposes.
{
\small
\begin{subequations}
    
    \begin{equation}
        \max \sum_{\substack{t \in \T \\ c,r \in \C \\ s \in \SD}}(\frac{g_{t,c,r,s}^{\text{DL}}}{D^{\text{DL}}} + \frac{g_{t,c,r,s}^{\text{UL}}}{D^{\text{UL}}}) \label{milp:of_mtf},  
    \end{equation}
    {\normalsize   
    \textit{Objective function (\ref{milp:of_mtf}):} the objective function of the MTF aims to maximize the sum-throughput of all UEs, considering both uplink and downlink. The overall uplink/downlink throughput is normalized by per-UE traffic requests in order to achieve a balanced uplink/downlink maximization.}

    \begin{flalign}
    &y_c^{\text{IAB}}+y_c^{\text{RIS}}+y_c^{\text{NCR}} \leq 1, &\forall c \in \C,\label{milp:one_tech}\\
    & y_{\hat{c}}^{\text{DON}}  \geq 1,\label{milp:don_res}\\
    & \sum_{c \in \C}y_c^{\text{DON}} = 1,\label{milp:one_don}\\
    &y_{\hat{c}}^{\text{DON}} \leq y_{\hat{c}}^{\text{ IAB}},\label{milp:don_prom}\\
    & y_{\Tilde{c}}^{\text{RIS}}  \geq 1,\label{milp:fake_res}
  
    \end{flalign}
    {\normalsize 
    \textit{Deployment Constraints (\ref{milp:one_tech}-\ref{milp:fake_res}):} Constraint~(\ref{milp:one_tech}) ensures mutual exclusivity among IAB Nodes, RISs, and NCRs in a specific CS $c \in \C$. Constraints (\ref{milp:don_res}) and (\ref{milp:one_don}) guarantee the presence of a single Donor in $\hat{c} \in \C$. Constraint~(\ref{milp:don_prom}) allows the promotion of the IAB Node installed in $\hat{c}$ to an IAB Donor. Constraint (\ref{milp:fake_res}) install a "fake" RIS \footnote{The model does not distinguish between an RIS or an NCR for this CS, as its purpose is just to indicate that no SRD must be installed .} $\Tilde{c}\in \C$.}
    
    \begin{flalign}
    &\sum_{\substack{c \in \C \setminus \\ \{\hat{c},\Tilde{c}\}}}\left( P_{c}^{\text{IAB}}y_c^{\text{IAB}} + P_{c}^{\text{RIS}}y_c^{\text{RIS}} + P_{c}^{\text{NCR}}y_c^{\text{NCR}} \right) \leq B&\label{milp:budget}
    \end{flalign}
    {\normalsize \textit{Budget Constraint:} Constraint~(\ref{milp:budget}) limits the installation budget to $B$. Note that the sum excludes the IAB Donor (as it represents a fixed cost) and the "fake" SRD.}

    \begin{flalign}
    &z_{c,d} \leq \Delta_{c,d}^{\text{BH}}(y_c^{\text{IAB}} + y_d^{\text{IAB}})/2, &\forall  c,d \in \C\label{milp:bh_act}\\
    &x_{t,c,r,RIS} \leq  \Delta_{t,c,r,RIS}^{\text{SRC}}(y_c^{\text{IAB}} + y_r^{\text{RIS}})/2,&\forall t \in \T, c,r \in \C\label{milp:src_ris_act}\\
    &x_{t,c,r,NCR} \leq  \Delta_{t,c,r,NCR}^{\text{SRC}}(y_c^{\text{IAB}} + y_r^{\text{NCR}})/2,&\forall t \in \T, c,r \in \C\label{milp:src_ncr_act}\\
    & \sum_{\substack{c,r \in \C,\\ s \in \SD}} x_{t,c,r,s} = 1, &\forall  t \in \T\label{milp:single_acc}\\
    &\sum_{d \in \C}z_{d,c} \leq 1 - y_c^{\text{DON}}, &\forall c \in \C\label{milp:tree}
    \end{flalign}
    \vspace{-0.4cm}
    \begin{flalign}
    &\sum_{s \in \SD}x_{t,c,r,s} \leq b_{c,r}, &\forall t \in \T, c,r \in \C,\label{milp:excl_def}\\
    &\sum_{c \in \C}b_{c,r} \leq 1, &\forall r \in \C\setminus{\Tilde{c}}\label{milp:excl_enforce}
    \end{flalign}
    {\normalsize
    \textit{Link-Activation Constraints (\ref{milp:bh_act}-\ref{milp:excl_enforce}):} Constr. (\ref{milp:bh_act}) permits the activation of a backhaul link connecting two CSs $c,d \in \C$ under the conditions that both CSs have an installed IAB Node and the link can be potentially established ($\Delta_{c,d}^{BH} = 1 $). Similarly, in (\ref{milp:src_ris_act}) the activation of an RIS-based SRC can occur when a BS is installed in $c$ and an RIS in $r$. In (\ref{milp:src_ncr_act}), an NCR-based SRC can be established when a BS is installed in $c$ and an NCR in $r$. Constr.  (\ref{milp:single_acc}) ensures that each TP is served by a unique SRC, while (\ref{milp:tree}) enforces a tree topology across backhaul links. 

    Constr. (\ref{milp:excl_def}) and Constr. (\ref{milp:excl_enforce}), relying on the auxiliary decision variable $b_{c,r}$, enforce that each SRD in CS $r$ must be associated with and controlled by a unique IAB device in CS $c$, and thus serving only the UEs of that IAB device. This is a practical constraint as SRDs operates under the master control of a RAN device.
}
    
    \begin{flalign}
    &   w_c^{\text{DL}} + \sum_{d \in \C} (f_{d,c}^{\text{DL}} - f_{c,d}^{\text{DL}}) - \sum_{\substack{t \in \T, \\ r \in \C, s \in \SD}} g_{t,c,r,s}^{\text{DL}}= 0 &\forall  c \in \C,\label{milp:flow_bal_dl}\\
    &    - w_c^{\text{UL}} + \sum_{d \in \C} (f_{d,c}^{\text{UL}} - f_{c,d}^{\text{UL}}) + \sum_{\substack{t \in \T, \\ r \in \C, s \in \SD}} g_{t,c,r,s}^{\text{UL}}= 0 &\forall  c \in \C,\label{milp:flow_bal_ul}
    \end{flalign}
    \vspace{-0.4cm}
    \begin{flalign}
    & f_{c,d}^{\text{DL}} \leq C_{c,d}^{\text{BH}} z_{c,d}, &\forall c,d \in \C,\label{milp:bh_cap_dl}\\
    & f_{d,c}^{\text{UL}} \leq C_{d,c}^{\text{BH}} z_{c,d}, &\forall c,d \in \C,\label{milp:bh_cap_ul}\\
    & g_{t,c,r,s}^{\text{DL}} \geq  D^{\text{DL}} x_{t,c,r,s}, &\forall t \in \T, c, r \in \C, s \in \SD,\label{milp:acc_min_demand_dl}\\
    & g_{t,c,r,s}^{\text{UL}} \geq  D^{\text{UL}} x_{t,c,r,s}, &\forall t \in \T, c, r \in \C,s \in \SD,\label{milp:acc_min_demand_ul}\\
    & g_{t,c,r,s}^{\text{DL}} \leq  C_{t,c,r,s}^{\text{SRC,DL}} x_{t,c,r,s}, &\forall t \in \T, c, r \in \C,s \in \SD,\label{milp:acc_cap_dl}\\
    & g_{t,c,r,s}^{\text{UL}} \leq C_{t,c,r,s}^{\text{SRC,UL}} x_{t,c,r,s}, &\forall t \in \T, c, r \in \C,s \in \SD,\label{milp:acc_cap_ul}\\
     & w_c^{\text{DL}} + w_c^{\text{UL}} \leq M y_c^{\text{DON}}, &\forall c \in \C\label{milp:donor_cap}
     \end{flalign}
    {\normalsize 
    
    \textit{Flow Constraints (\ref{milp:flow_bal_dl}-\ref{milp:donor_cap}):} Constraint (\ref{milp:flow_bal_dl}) and Constraint (\ref{milp:flow_bal_ul}) ensure flow balance at any device in the backhaul tree for both downlink and uplink flows. Constraints (\ref{milp:bh_cap_dl}) and (\ref{milp:bh_cap_ul}) set upper bounds on backhaul link flows to match link capacities. Constraints (\ref{milp:acc_min_demand_dl}) and (\ref{milp:acc_min_demand_ul}) guarantee minimum per-UE traffic requests $(D^{\text{DL}},D^{\text{UL}})$, while Constraints (\ref{milp:acc_cap_dl}) and (\ref{milp:acc_cap_ul})  set a maximum capacity for SRC-based access links. Finally, Constraint (\ref{milp:donor_cap}) restricts the traffic exchanged between the RAN and the Core Network, assuming it cannot exceed the capacity, denoted by $M$\footnote{$M$ is the maximum value among the averages of the downlink/uplink capacities on all Donor links, weighted by the selected TDD downlink/uplink split ratio.}, of the best Donor link.}
 
    \begin{flalign}
    &\sum_{d \in \C} (\frac{f_{c,d}^{\text{DL}}}{C_{c,d}^{\text{BH}}} + \frac{f_{d,c}^{\text{DL}}}{C_{d,c}^{\text{BH}}}) + \sum_{\substack{t \in \T, \\ r \in \C,  s \in \SD}}\frac{g^{\text{DL}}_{t,c,r,s}}{C_{t,c,r,s}^{\text{SRC,DL}}}=t_c^{\text{DL}},&\forall c \in \C,\label{milp:dl_ratio}\\
    &\sum_{d \in \C} (\frac{f_{c,d}^{\text{UL}}}{C_{c,d}^{\text{BH}}} + \frac{f_{d,c}^{\text{UL}}}{C_{d,c}^{\text{BH}}}) + \sum_{\substack{t \in \T, \\ r \in \C,  s \in \SD}}\frac{g^{\text{UL}}_{t,c,r,s}}{C_{t,c,r,s}^{\text{SRC,UL}}}=t_c^{\text{UL}},&\forall c \in \C\label{milp:ul_ratio}\\
    &t_c^{\text{DL}}\leq\alpha y_c^{\text{IAB}},&\forall c \in \C,\label{milp:dl_tdd}\\
    &t_c^{\text{UL}}\leq(1-\alpha)y_c^{\text{IAB}},&\forall c \in \C\label{milp:ul_tdd}
    \end{flalign}
    {\normalsize 
    \textit{Resource-sharing Constraints (\ref{milp:dl_ratio}-\ref{milp:ul_tdd}):} Constraint (\ref{milp:dl_ratio}) implements the concept of time-sharing. We assume $t_c^{DL} \in [0,1]$ to represent the time share available for downlink transmissions (along access and backhaul links) at the IAB device in CS $c$. Indeed, the time share each link must work to support a given flow is the ratio between the flow intensity and the nominal capacity of the link. Constraint (\ref{milp:ul_ratio}) has the same meaning for uplink transmissions. 
The TDD frame split is enforced by Constraints (\ref{milp:dl_tdd}) and (\ref{milp:ul_tdd}). Downlink resources can utilize up to $\alpha (<1)$ fraction of the total device's resources, and consequently, uplink can occupy no more than $(1 - \alpha)$ of the total resources. Note that these constraints not only model the uplink/downlink TDD frame split (e.g., the 4:1 DL/UL scheme becomes $\alpha=\frac{4}{5}$) but also enforce half-duplex operations by considering both transmitted and received flows.}

    \begin{flalign}
    &\phi_r \geq \Phi_{r,t}^{\text{A}}-\frac{F}{2} - 2\pi(1-\sum_{s\in\SD}x_{t,c,r,s}),\hspace{-1em}&&\forall t \in \T, c,r \in \C\setminus {\Tilde{c}},\label{milp:min_ang_tp}\\
    &\phi_r \leq \Phi_{r,t}^{\text{A}}+\frac{F}{2} + 2\pi(1-\sum_{s\in\SD}x_{t,c,r,s}),\hspace{-1em}&&\forall t \in \T, c,r \in \C\setminus {\Tilde{c}},\label{milp:max_ang_tp}\\
    &\phi_r \geq \Phi_{r,c}^{\text{B}}-\frac{F}{2} - 2\pi(1-x_{t,c,r,RIS}),\hspace{-1em}&&\forall t \in \T, c,r \in \C\setminus {\Tilde{c}},\label{milp:min_ang_ris}\\
    &\phi_r \leq \Phi_{r,c}^{\text{B}}+ \frac{F}{2} + 2\pi(1-x_{t,c,r,RIS}),\hspace{-1em}&&\forall t \in \T, c,r \in \C\setminus {\Tilde{c}},\label{milp:max_ang_ris}\\
    &\phi_r \geq \Phi_{r,c}^{\text{B}}+\frac{\pi}{2} - 2\pi(1-x_{t,c,r,NCR}),\hspace{-1em}&&\forall t \in \T, c,r \in \C\setminus {\Tilde{c}},\label{milp:min_ang_ncr}\\
    &\phi_r \leq \Phi_{r,c}^{\text{B}}- \frac{\pi}{2} + 2\pi(1-x_{t,c,r,NCR}),\hspace{-1em}&&\forall t \in \T, c,r \in \C\setminus {\Tilde{c}}\label{milp:max_ang_ncr}
\end{flalign}
{\normalsize 
\textit{SRD-orientation Constraints. (\ref{milp:min_ang_tp}-\ref{milp:max_ang_ncr}):} These constraints determine the value of the SRD rotation variable $\phi_r$ based on the angles between the involved devices\footnote{Set element $\Tilde{c}$ is not considered in this constraint set since the "fake" SRD must not be limited in orientation as it doesn't exist.}.

An RIS consists of a single panel (surface) with a limited (FoV) $F$. Therefore, the orientation of an RIS must ensure that both directions connecting the RIS to the IAB node and to the UE fall within the RIS's FoV. 

An NCR comprises two adjustable panels, each characterized by its FoV. One panel of the NCR must be oriented toward the IAB Node (BS panel), which we assume can be perfectly pointed. The other panel is directed toward the UEs (UE panel). However, the two panels are not freely adjustable, and a minimum angle of $\frac{\pi}{2}$ between their directions is necessary for effective interference cancellation techniques, ensuring sufficient isolation for simultaneous transmission and reception. 

Finally, note that an RIS or a UE panel can be involved in multiple SRC instances providing access to several UEs. Therefore, the LoS to all these UEs must lie within the panel's FoV.

The rotation variable $\phi_r \in [0,2\pi]$ denotes the orientation (relative to a reference direction) of the SRD installed in CS $r$. If the SRD is an RIS, $\phi_r$ represents the orientation of the normal to its surface. If the SRD is an NCR, $\phi_r$ indicates the orientation of the normal of the UE panel, as no choice needs to be made for the  BS panel perfectly pointed toward the IAB Node.

Constraints (\ref{milp:min_ang_tp}) and (\ref{milp:max_ang_tp}) ensure that the (SRD - UE) link falls in the FoV of the oriented  RIS surface or UE panel. If the SRD is an RIS, the (IAB Node - RIS) link must lie in the same FoV as well, as enforced by Constraints (\ref{milp:min_ang_ris}) and (\ref{milp:max_ang_ris}). The minimum angular separation of $\frac{\pi}{2}$ between BS and UE panels of an NCR  is enforced  by (\ref{milp:min_ang_ncr}) and (\ref{milp:max_ang_ncr}).}
\end{subequations}
}

\subsection{Peak-Throughput Formulation (PTF)}\label{sub:ptf}
In this formulation, we want to model the bursty nature of the mmWave traffic. Given an achievable peak rate much higher than mean rate, mmWave data transfers will be very short, working at Gbps. In this context, the probability that two short bursts occur at the same time can be reasonably considered negligible, therefore the peak throughput of each burst can be assumed to individually occupy the entire set of resources left unused by the mean-throughput traffic, in practice, as it was isolated from other bursts. 

The Peak-Throughput Formulation (PTF) extends the Mean-Throughput Formulation (MTF) by introducing additional variables and constraints. Indeed, the allocation of peak-throughput traffic requires a modeling approach different from the one of the mean-throughput traffic in MTF, where every UE shares all available resources.
Additional flow variables, $f_{t,c,d}^{\text{X,UL}}$ and $f_{t,c,d}^{\text{X,DL}}$, are introduced to capture the peak-throughput traffic in the wireless backhaul, while  $g_{t,c,r,s}^\text{X,DL}$ and $g_{t,c,r,s}^\text{X,UL}$ represent peak-throughput access traffic . The mean-throughput traffic, which is maximized in MTF, becomes a constraint for PTF, which considers minimum mean-throughput traffic guarantees (expressed by $D^{\text{DL}}$ and $D^{\text{UL}}$) that partially occupy the network capacity. 
The rationale behind PTF is to optimize the mean-throughput traffic flow distribution to increase the capacity of the bottlenecks that will be saturated as soon as peak-rate bursts start.

The objective function and constraints characterizing PTF are:
{
{
\small
\begin{subequations}
\begin{equation}
    \max \sum_{\substack{t \in \T,\\c,r \in \C,s \in \SD}}\frac{g_{\substack{t,c,\\r,s}}^{\text{X,DL}}}{D^{\text{DL}}} + \frac{g_{\substack{t,c,\\r,s}}^{\text{X,UL}}}{D^{\text{UL}}},\label{milp:of_ptf}
\end{equation}
{\normalsize \textit{Objective function:} The objective function for the PTF, as expressed in (\ref{milp:of_ptf}), aims to maximize the sum of users' peak throughputs considering both uplink and downlink. These values are  normalized as in the MTF's objective function.}
\begin{flalign}
&   w_{t,c}^{\text{X,DL}} + \sum_{d \in \C} (f_{t,d,c}^{\text{X,DL}} - f_{t,c,d}^{\text{X,DL}}) - \sum_{\substack{r \in \C,\\s\in\SD}} g_{\substack{t,c,\\r,s}}^{\text{X,DL}}= 0 &\forall  c \in \C,t \in \T,\label{milp:x_flow_bal_dl}\\
&   - w_{t,c}^{\text{X,UL}} + \sum_{d \in \C} (f_{t,d,c}^{\text{X,UL}} - f_{t,c,d}^{\text{X,UL}}) + \sum_{\substack{r \in \C,\\s\in\SD}} g_{\substack{t,c,\\r,s}}^{\text{X,UL}}= 0 &\forall  c \in \C,t \in \T,\label{milp:x_flow_bal_ul}
\end{flalign}
\vspace{-0.4cm}
\begin{flalign}
& f_{t,c,d}^{\text{X,DL}} \leq C_{c,d}^{\text{BH}} z_{c,d}, &\forall c,d \in \C, t \in \T,\label{milp:x_bh_cap_dl}\\
& f_{t,d,c}^{\text{X,UL}} \leq C_{d,c}^{\text{BH}} z_{c,d}, &\forall c,d \in \C, t \in \T,\label{milp:x_bh_cap_ul}\\
& g_{\substack{t,c,\\r,s}}^{\text{X,DL}} \leq C_{\substack{t,c,\\r,s}}^{\text{SRC,DL}} x_{t,c,r,s}, &\forall t \in \T, c, r \in \C,s \in \SD,\label{milp:x_acc_cap_dl}\\
& g_{\substack{t,c,\\r,s}}^{\text{X,UL}} \leq C_{\substack{t,c,\\r,s}}^{\text{SRC,UL}} x_{t,c,r,s}, &\forall t \in \T, c, r \in \C,s \in \SD,\label{milp:x_acc_cap_ul}\\
& w_{t,c}^{\text{X,DL}} + w_{t,c}^{\text{X,UL}} \leq M y_c^{\text{DON}}, &\forall t \in \T, c \in \C\label{milp:x_donor_cap}
\end{flalign}
{\normalsize \textit{Peak-throughput flow constraints (\ref{milp:x_flow_bal_dl}-\ref{milp:x_donor_cap}):} 
Similar to the corresponding constraint group in MTF, constr. (\ref{milp:x_flow_bal_dl}) and constr. (\ref{milp:x_flow_bal_ul}) ensure flow balance at any device in the backhaul  tree for both uplink and downlink peak flows. Constr. (\ref{milp:x_bh_cap_dl}) and constr. (\ref{milp:x_bh_cap_ul}) set upper bounds on backhaul peak flows according to link capacities, while constr. (\ref{milp:x_acc_cap_dl}) and constr. (\ref{milp:x_acc_cap_ul}) impose limits on the maximum SRC access throughput. Finally, constr. (\ref{milp:x_donor_cap}) restricts the traffic between the RAN and the Core Network.
} 

\begin{flalign}
&\sum_{d \in \C} (\frac{f_{t,c,d}^{\text{X,DL}}}{C_{c,d}^{\text{BH}}} + \frac{f_{t,d,c}^{\text{X,DL}}}{C_{d,c}^{\text{BH}}})+\nonumber\\ 
&+ \sum_{\substack{r \in \C,\\s\in\SD}}\frac{g_{\substack{t,c,\\r,s}}^{\text{X,DL}}}{C_{\substack{t,c,\\r,s}}^{\text{SRC,DL}}}\leq\alpha y_c^{\text{IAB}}-t_c^{\text{DL}},&\forall c \in \C, t \in \T\label{milp:x_tdd_dl},\\
&\sum_{d \in \C} (\frac{f_{t,c,d}^{\text{X,UL}}}{C_{c,d}^{\text{BH}}} + \frac{f_{t,d,c}^{\text{X,UL}}}{C_{d,c}^{\text{BH}}}) + \nonumber\\
&+\sum_{\substack{r \in \C,\\s\in\SD}}\frac{g_{\substack{t,c,\\r,s}}^{\text{X,UL}}}{C_{\substack{t,c,\\r,s}}^{\text{SRC,UL}}}\leq(1-\alpha)y_c^{\text{IAB}}-t_c^{\text{UL}},&\forall c \in \C,t \in \T\label{milp:x_tdd_ul},
\end{flalign}
\end{subequations}

{\normalsize \textit{Peak-throughput resource-sharing constraints (\ref{milp:x_tdd_dl}-\ref{milp:x_tdd_ul}):}
Constraints (\ref{milp:x_tdd_dl})-(\ref{milp:x_tdd_ul}) are equivalent to constraints (\ref{milp:dl_tdd})-(\ref{milp:ul_tdd}) in Sec.~\ref{sub:mtf}: the peak-throughput additional traffic associated with TP $t\in\T$ must be regulated by the capacity of the bottleneck BS on the path to the IAB Donor, which takes into account the average traffic ($D^{\text{UL}}$ and $D^{\text{DL}}$ ) to be guaranteed in the TDM frame (captured by time fractions $t_c^{\text{DL}}$ and $t_c^{\text{UL}}$).}
 }
}

\subsection{Peak-Throughput Heuristics}\label{sub:fullmodel_heur}
The increased complexity introduced by the PTF required an MILP model that proved challenging to solve in a time window of 60 minutes. Consequently, in order to tackle larger instances and analyze many instances for each parameter setting, we have implemented a heuristic approach to speed up the solution process while maintaining high-quality solutions. {\color{black} The main steps of the heuristic approach are summarized in Algorithm \ref{alg:ptf-heu}}. In this context, two changes have been applied to the MILP solution process. {\color{black} In Step 1}, we have applied a technique, which we introduced in ~\cite{10293769}, that consists in reducing the parameter $M$, the capacity of the best backhaul link of the IAB Donor, \textcolor{black} { to $M^{'}=M/10$.} {\color{black} We found that scaling the parameter to a fraction of its actual value, thus creating a bottleneck effect, reduces the solution time to just a few seconds, while keeping the upper-bounded optimality gap (\textit{UB-optgap}, i.e., the difference between the best feasible solution and the continuous relaxation upper bound) below 5\%. In Step~2, we relaxed the UB-optgap tolerance from $5\%$ (used in MTF) to $40\%$.} This adjustment is motivated by the behavior of the solution exploration process, which  identifies feasible solutions in the first seconds and, most of the times, spends the remaining time in improving the continuous relaxation, without remarkably increasing the quality of found solutions. {\color{black} It is also worth noting that the UB-optgap of 40\% occurs only in a few, unfavorable instances; on average, the UB-optgap is about 22\% when the optimizer is stopped after 30 minutes of execution.}  
While this adjustment theoretically provides a lower bound of the exact optimal solution, the results presented in Section~\ref{sec:results} indicate a large PTF performance improvement compared to MTF.

{
\color{black}

\begin{algorithm}
   \caption{PTF Heuristic Algorithm}\label{alg:ptf-heu}
\begin{algorithmic}[1]
\color{black}
    \State \texttt{\textbf{M' $\gets$ M / 10}} \Comment{\footnotesize The value of $M$ is reduced to $M'$ to shrink the solution space by creating an artificial bottleneck};
    \State \texttt{\textbf{relmipgap $\gets$ 0.4}} \Comment{\footnotesize Relaxing optimality gap shortens execution times by reducing the effort to find tight optimality gap guarantees};
    \State \texttt{\textbf{Instance loading}} \Comment{\footnotesize Files including models in \S \ref{sub:mtf}, parameters in Table~\ref{tab:pars}, and solution meta-parameters are sent to the solver};
    \State \texttt{\textbf{Solution}} \Comment{\footnotesize The solver runs the optimization process and returns the obtained solution}.
\end{algorithmic}
\end{algorithm}
}

{\color{black}
\subsection{Other traffic types}\label{sub:other_ traffic}
The Ericsson Mobility Report \cite{ericssonReportJun2025} shows that about 80\% of mobile network traffic consists of video, social networking, communication apps, cloud storage, software downloads, and file sharing, all of which are well represented by our mean and peak throughput model. Nevertheless, other traffic categories, such as online gaming or mission-critical applications, may require QoS guarantees like low latency or a guaranteed bitrate. These requirements can be incorporated into our framework with minimal changes to the formulation.

For guaranteed bitrate, both MTF and PTF already enforce UL and DL demands, $D^{UL}$ and $D^{DL}$, while optimizing mean and peak throughput. Additional guarantees can be included by extending these as
\[
D^{x,QoS} = D^{x} + R_{GBR}^{x}, \quad x \in \{\text{UL}, \text{DL}\},
\]
where $R_{GBR}^{x}$ is the guaranteed bitrate.

Latency constraints can also be addressed at the planning level. While latency-sensitive traffic is mostly handled at the gNB through scheduling, bearers, and semi-persistent allocation, the planning stage must reserve capacity to support these mechanisms. This can be done by applying an overload factor $\alpha$ to the demands,
\[
D^{x,QoS} = D^{x}(1+\alpha), \quad x \in \{\text{UL}, \text{DL}\},
\]
or by scaling down capacities with a spare-capacity factor $\beta$, e.g.,
\[
C_{c,d}^{\text{BH},QoS} = \beta \cdot C_{c,d}^{\text{BH}},
\]
with typical $\beta$ values in $[0.7, 0.8]$. Reserving spare capacity is a standard practice to mitigate latency issues.

With these QoS-aware parameters, both MTF and PTF can be solved directly with only minor modifications, without altering the overall approach or conclusions.

}

\section{Results}
\label{sec:results}
In this section, we analyze the optimal solutions provided by the MILP solver for the two formulations presented in Section~\ref{sec:opt_models}. We aim at comparing the outcomes of the two optimization approaches in order to understand the key features of resulting topologies and SRD deployment strategies.

The set of 100 considered instances are based on hexagonal cell areas with a radius of $150$ meters, located a random points in a map (including streets and 3D building polyhedra) of the Milan urban area~\cite{MilanoGeoportale}. Instances have been selected in a heterogeneous manner, considering the diversity of building clutter, to ensure a representation of both open spaces (such as squares and parks) and urban canyons (like densely populated residential areas). Indeed, this approach allows to evaluate the impact of a realistic building distribution characterizing a typical European metropolitan city. 

Each cell has an IAB Donor located at the leftmost vertex of the hexagonal area, while set cardinalities $|\C|=25$, $|\T|=15$ and $|\SD|=2$ represent CSs, TPs and types of SRDs, respectively. The coordinates of the elements of these sets are randomly selected in the cell area, avoiding the space inside buildings.

\begin{figure}
    \centering
    \captionsetup{justification=centering,margin=1.5cm}
    \vspace{-0.7 cm}
    \includegraphics[width=0.8\columnwidth]{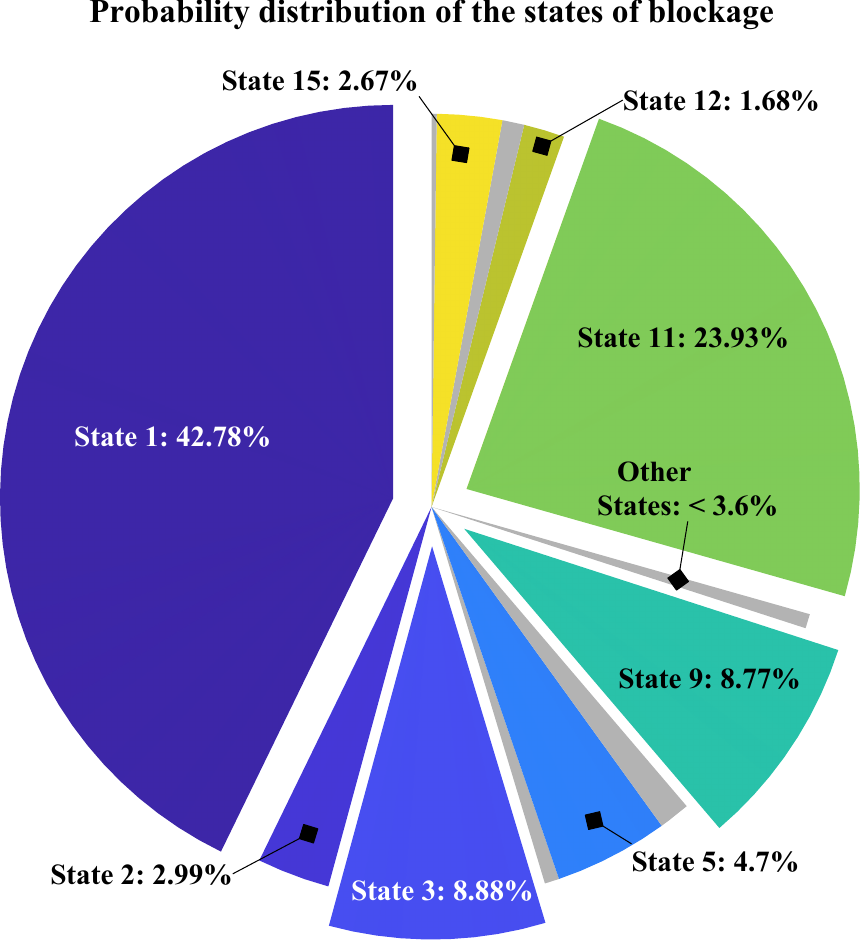}
    \caption{Probability distribution of the blockage states \textcolor{black}{from Table~\ref{tab:16_states}}. The largest probabilities are indicated.}
    \label{fig:state_pie}
\end{figure}

IAB Nodes and IAB Donor are equipped with three $120^{\circ}$ sectors each consisting, respectively, of a $16\times12$-element and $12\times8$-element panel array with $58$ dBm and $51$ dBm of Effective Isotropic Radiated Power (EIRP), a carrier frequency of $28$ GHz and a bandwidth of $400$ MHz. UEs are modeled as $2\times 2$ antenna arrays with an EIRP of $29$ dBm, while RISs consist of $100\times100$ passive elements in a rectangular array, with a FoV of $170^{\circ}$. NCRs are made of two $12\times6$ panels with an EIRP of $50$ dBm. All devices have $\frac{\lambda}{2}$ spacing between elements in both directions. The IAB Donor is placed at a $25$ m of height, the IAB Nodes at $6$ m, the RISs and NCRs at $3$ m, and the UEs at $1.5$ m. The cost of an IAB Node is set as a reference $1$ unit. RISs, less expensive than IAB Nodes, can be assumed to reach a commercial price ten times lower, therefore a normalized cost of $0.1$ units. NCRs can be set to an intermediate price of $0.5$ units, due to their amplifying capabilities. 

A first result is shown in Figure~\ref{fig:state_pie}: it represents the probability of finding an SRC in each of the 16 blockage states of Table~\ref{tab:16_states}, evaluated over each potential SRC triplet -- IAB node, SRD, and UE -- of all 100 urban cells. Note that this distribution can vary according to the parameters defining blockage probabilities, such as obstacle size and density, the probability of holding the device in a specific orientation, and the SBR angle. In this study, we consider a density of $0.002$ obst/$\text{m}^2$ and nomadic obstacles with a size of $1.6\times1.8\times4.5$ $m^3$, an equal probability of holding the UE in either portrait or landscape mode ($P_p=P_l=0.5$), and SBR angles consistent with 3GPP Table~\ref{tab:3GPP_table}. We can notice that states 1, 11, 3, and 9 collectively represent more than the 80\% of the total. These states correspond to scenarios where both links are not blocked~(1), both links experience self-blockage~(11), only the direct link is in the SBR~(3), or only the indirect link is in the SBR~(9). Therefore, self-blockage emerges as the predominant blockage cause.  Indeed, nomadic obstacles, which have the same size as an average vehicle, impact a wireless link less than 20\% of the time.

Two assessment campaigns have been conducted: one varying the available budget for the deployment (Sec.~\ref{sub:budget}) and the other changing the minimum demand guaranteed to the UEs (Sec.~\ref{sub:demand}). These campaigns investigate the main features characterizing network planning: device installation, throughput, topological features, and bottleneck distribution. The total available budget spans from $0$ to $12$ units, while the guaranteed minimum overall UE demand varies from  $2.5$ to $250$ Mb/s (to be split 4:1 between downlink and uplink, reflecting TDD frame split). All the link capacities for SRCs are computed according to Eq.~(\ref{eq:3gpp_cap_huawei}), and Eq.~(\ref{eq:src_cap}). All the results are averaged over the $100$ cell instances generated through MATLAB and solved by IBM ILOG CPLEX optimizer. The UB-optgap is set at $5\%$ for MTF, while the peak-throughput heuristic of Sec.~\ref{sub:fullmodel_heur} is used for PTF. All the processing has been carried out on a $40$-core Intel Xeon E5-2640 v4 @ $2.40$GHz machine with $128$ GB of RAM.

{\color{black} Although solving a single network planning instance in several hours, or even days, is common in real scenarios, to capture meaningful trends within reasonable execution times we solved 100 random instances with a 30-minute solver limit per instance for each point in the following plots.

Since longer runtimes can reduce the optimality gap or enable the solution of larger instances, we performed a scalability analysis that showed problems with 40 Candidate Sites (CSs) and 24 Test Points (TPs) can be solved within 2 hours with comparable gaps. As these scenarios, which are the object of the optimization, correspond to the size of a typical urban macro cell, the number of potential sites considered is consistent with realistic deployments. Therefore, solving such NP-hard instances (which can be reduced to a set cover problem) within a few hours is computationally feasible and fully compatible with practical network planning.}

\subsection{Budget campaign}
\label{sub:budget}
This campaign aims to assess the impact of the deployment budget on the optimal network design and its performance. In order to carry out the comparison, we set a minimum guaranteed user demand to $120$ Mb/s in downlink and $30$ Mb/s in uplink. Throughout the figures of this campaign, a black vertical line at budget value equal to $10$ highlights the data intersection with the next campaign in Sec.~\ref{sub:demand}.

A first experimental outcome is that no solution can be found when the available budget is less than $1$ unit: the IAB Donor without the support of at least one IAB Node is not enough to guarantee the minimum throughput requirements.

\begin{figure}
    \centering \captionsetup{justification=centering,margin=1.5cm}  
    \includegraphics[width=0.70\columnwidth]{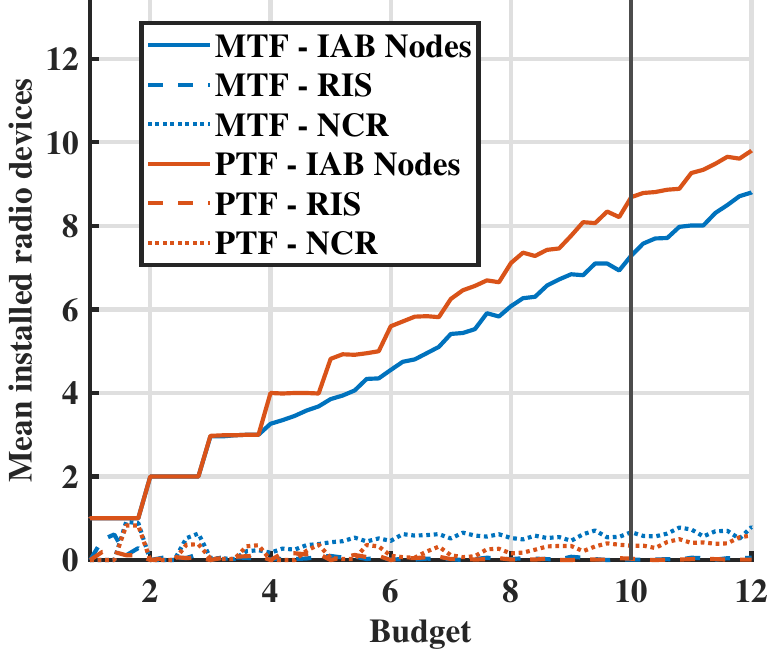}
    \caption{Number of different installed devices when varying the budget. \textcolor{black}{In both formulations, the expenditure is dominated by IAB Nodes.}}
    \label{fig:dev_b}
\end{figure}

Figure~\ref{fig:dev_b} shows the variation in the number of installed device numbers and their types when increasing the available budget for both MTF and PTF. The curves of the two formulations show a similar behavior: more IAB nodes are installed as the budget increases. Indeed, IAB nodes are very effective in providing high mean and peak throughputs to UEs. 
RISs and NCRs, instead, experience limited usage. There are few instances with deployed RISs, and an NCR is installed approximately every other instance. The reason behind this behavior is related to the throughput maximization objective, which implies a maximization of the (mean or peak) access capacity provided to each UE. The access capacity of an SRC consists in the average of the capacities experienced over all its blockage states (see Eq.~\ref{eq:src_cap}), however, although available only 50\% of the time, the rate achievable via a direct mmWave link is so large compared to the one obtainable via a reflection that is a dominant factor in the average. Therefore, the impact of RISs and NCRs on throughput maximization is minimal. RISs should not be used only to provide a high-rate alternative to the direct link. NCRs, due to their amplification capabilities, can support higher rates than RISs, therefore they are used more frequently.

\begin{figure}
\centering
\subfloat[Mean throughputs]{\includegraphics[width=0.45\columnwidth]{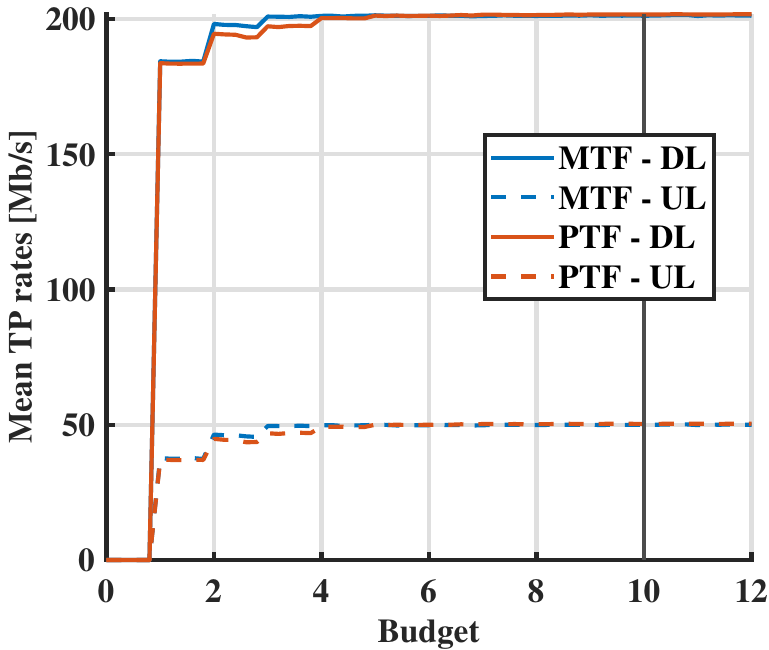}
\label{fig:mean_b}}
\hfil
\subfloat[Peak throughputs]{\includegraphics[width=0.45\columnwidth]{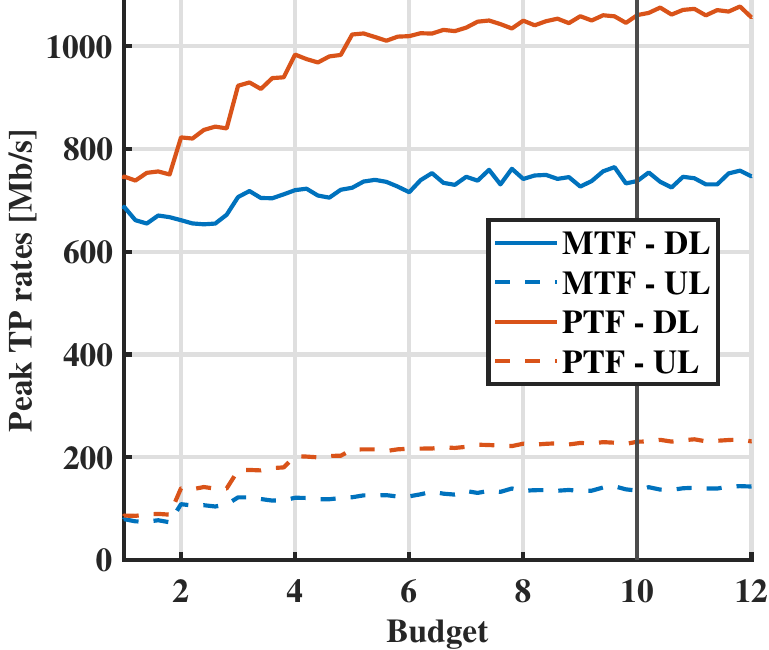}
\label{fig:peak_b}}
\caption{Overall throughput trends when budget varies (minimum guarantees: 120 Mb/s in
downlink and 30 Mb/s in uplink). \textcolor{black}{On the left (a), both formulations achieve similar mean throughputs per TP. On the right (b), PTF overwhelmingly exceeds MTF in peak throughput by 45\% (70\%) in DL (UL).}}
\label{fig:rates_b}
\end{figure}

Figures~\ref{fig:mean_b} and~\ref{fig:peak_b} present the user throughputs obtained in the networks designed according to the two formulations. In each network, we measure both the mean and the peak user throughput. Given that each formulation optimizes one of two metrics, it is important to appropriately handle the other metric to ensure a valid comparison of the results obtained from the two formulations. Specifically, for MTF, which optimizes the average throughput, the peak per-user throughput is determined by identifying the spare capacity on the bottleneck link for each specific user. For PTF, which optimizes the peak throughput, the mean throughput is computed by running an additional optimization using the MTF model, where only optimal routing and resource allocation are optimized within the network topology resulting from PTF. This approach results in instances evaluated not only in terms of the maximization objective of their own formulation but also according to the objective of the other formulation. A notable observation is that, despite PTF not specifically optimizing the mean throughput, it exhibits results remarkably similar to MTF ($200$ Mb/s in downlink and $50$ Mb/s in uplink). Nevertheless, PTF significantly outperforms MTF in terms of achievable peak throughput, with an improvement of up to $300$ Mb/s in downlink and $90$ Mb/s in uplink. Finally, all throughput curves show a saturation plateau after a specific budget level. This is caused by the the presence of a unique connection to the wired network, e.g., the IAB Donor, which is source and destination of all traffic flows and inherently presents a limited capacity of its wireless connections. When this limit is reached, deploying additional devices does not bring significant user throughput improvements.

\begin{figure}[H]
\centering
\subfloat[Number of hops from IAB Donor to UE]{\includegraphics[width=0.45\columnwidth]{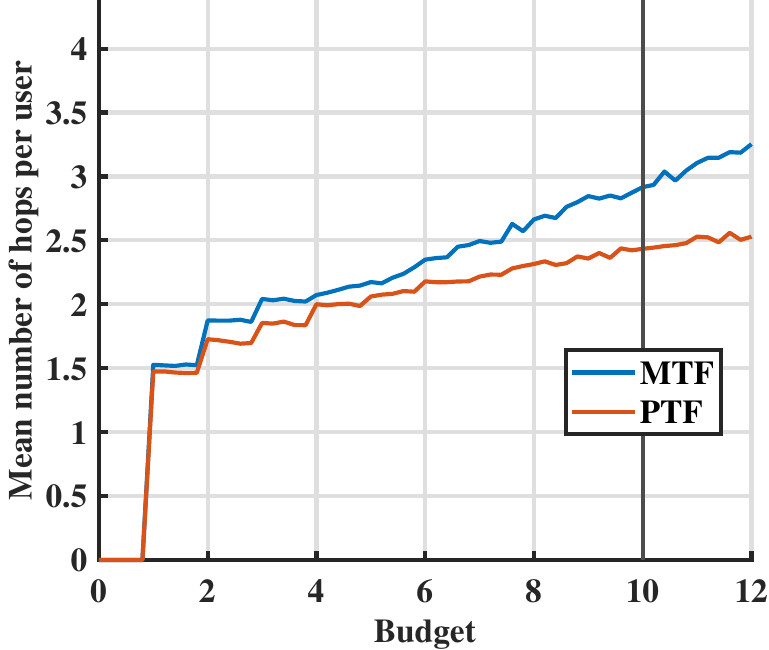}
\label{fig:mh_b}}
\hfil
\subfloat[Donor's node degree]{\includegraphics[width=0.45\columnwidth]{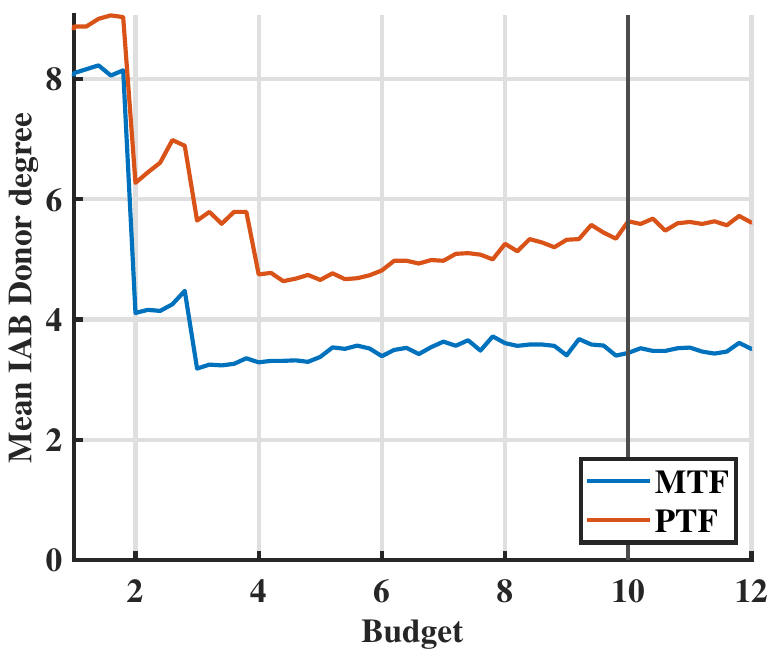}
\label{fig:dg_b}}
\caption{Variation of some topological properties over budget increase. \textcolor{black}{In (a), MTF generates longer hop chains compared to PTF; in (b), the number of outgoing links from the Donor is larger in PTF. This indicates a tendency toward star-like topologies when peak throughput is prioritized.}}

\label{fig:topo_b}
\end{figure}

The two formulations aim at the user throughput maximization from two different perspectives. The MTF's goal is to increase the average throughput of the entire set of concurrent user flows under a minimum guaranteed per-user throughput. PTF, instead, designs a layout that provides the same minimum guaranteed throughput to the concurrent user flows, but leaves a large spare capacity in each link so that individual traffic bursts can be accommodated at the highest rate possible. These two objectives are pursued at equal deployment budget. The two perspectives result into different topological choices, which emerge from Figure~\ref{fig:mh_b}, indicating the number of hops from the IAB Donor to each UE in downlink and uplink, and from Figure~\ref{fig:dg_b}, showing the IAB Donor's node degree, i.e., the number of links incident to the IAB Donor.

With a small budget, the IAB backhaul is very limited and the traffic is mainly collected by direct connections to the IAB Donor. This leads to few backhaul hops and a high IAB Donor's degree. However, as the budget increases, the access burden shifts towards IAB Nodes. This means larger backhaul networks and few connections departing from the IAB Donor towards IAB Nodes, which, in turn, connect to UEs.
The difference between MTF and PTF network layouts lies in the offset between these curves. MTF tends to form deeper wireless backhaul networks than PTF, characterized by a larger number of hops from the IAB Donor to each UEs. At the same time, the IAB Donor tends to have fewer incident links in MTF than in PTF. These metrics reveal how a peak-throughput optimization requires star-like topologies to reduce potential bottleneck points. {\color{black} Since the relaying devices are half-duplex, their capacity is inherently constrained and further reduced when forwarding traffic for multiple flows; adopting a star topology, where nodes connect directly to the donor, mitigates these limitations by avoiding multi-hop relaying and thereby alleviating bottlenecks.} In contrast, mean-throughput can be maximized with a properly-designed deeper backhaul, with less bottleneck concern.

\begin{figure*}[!ht]
    \centering   \captionsetup{justification=centering,margin=1.5cm}
    \includegraphics[width=0.80\textwidth]{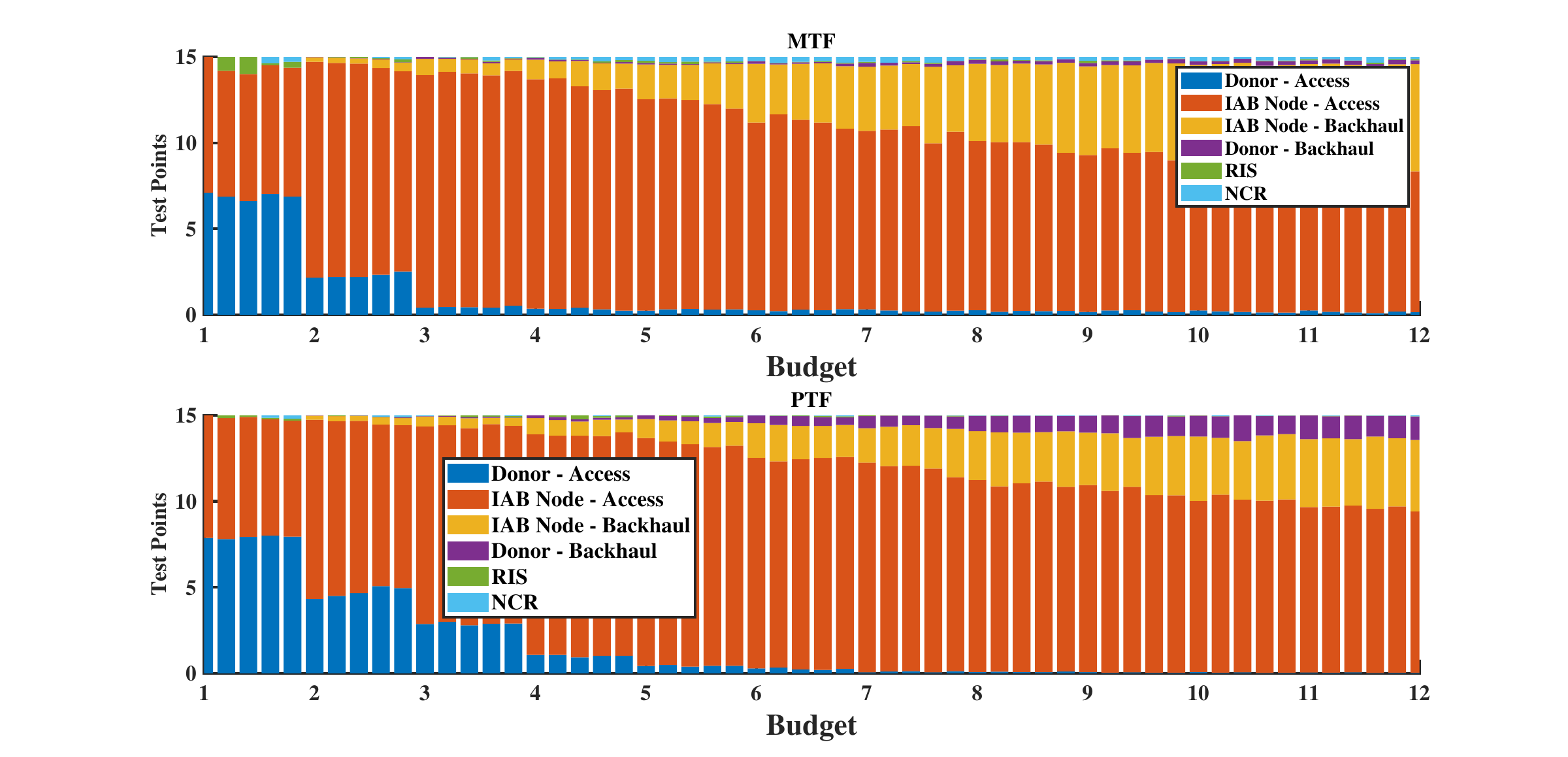}
    \caption{Distribution of bottlenecks over budget variation. \textcolor{black}{In both formulations, similar trends are observed at lower budgets: the main bottlenecks are access links through the IAB Donor or IAB Nodes. For PTF, given the star-like topology, the IAB Donor as a backhaul node becomes a more prominent bottleneck at higher budgets, as the network saturates.}}
    \label{fig:bn_b}
\end{figure*}
Figure~\ref{fig:bn_b} helps to shed some light on bottleneck causes. The figure indicates which device or link acts as a bottleneck for each user flow. At small budgets, the topology mainly consists of a single IAB Donor and few IAB nodes, therefore most UEs see a flow cap imposed by the saturation of IAB Donor's and IAB Nodes' access resources. When the budget increases more IAB nodes are installed, therefore the bottleneck distribution changes: only IAB Nodes' access links become the predominant source of bottlenecks.
Finally, from mid to high budgets, notice how the PTF formulation can design a backhaul that leads to the saturation of IAB Donor's backhaul links.

\subsection{Minimum demand campaign}
\label{sub:demand}
In this section, we discuss the results of a campaign focusing on the impact of the minimum guaranteed per-user demand. We set the deployment budget to a value of $10$ units and we vary the overall minimum guaranteed demand from $0$ to $250$ Mb/s, which according to the TDD split consists of $\frac{4}{5}$ in downlink and $\frac{1}{5}$ in uplink. Similarly to the previous figures, the vertical black line at the value of $150$ Mb/s serves as a reference point, marking the intersection with the data from the campaign in subsection~\ref{sub:budget}.

\begin{figure}
    \centering \captionsetup{justification=centering,margin=1.5cm}  \includegraphics[width=0.7\columnwidth]{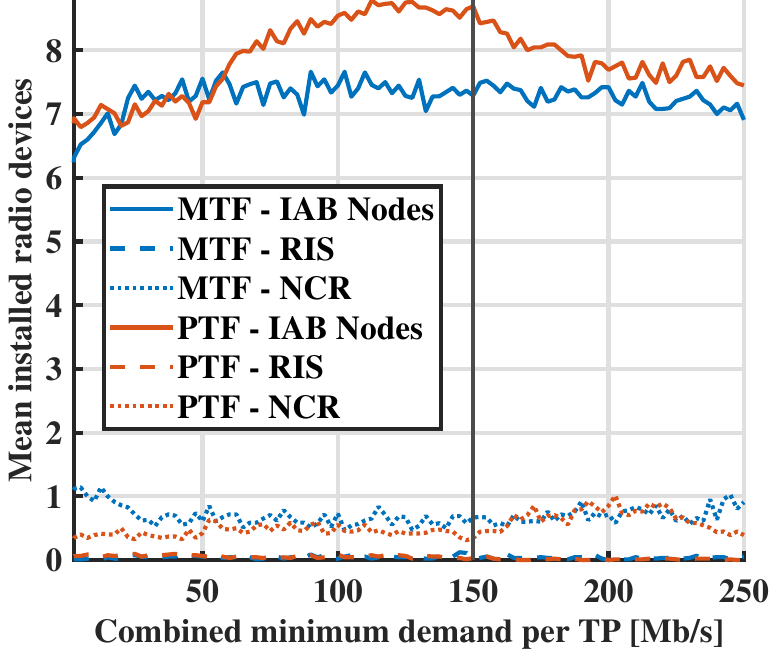}
    \caption{Number of different devices installed over demand variation. \textcolor{black}{As in the previous campaign, the most prominent device type are IAB Nodes. MTF maintains a quasi-constant trend of employed nodes throughout demand increase, while PTF deploys them more at medium demands.}}
    \label{fig:dev_d}
\end{figure}
Figure~\ref{fig:dev_d} shows the number of installed devices as the overall minimum  guaranteed demand increases, revealing the following trends:
\begin{enumerate*}[label={\textit{(\roman*)}}]

\item With large minimum demands ($200-250$ Mb/s), the lack of spare capacity for a peak throughput leads PTF to behave similarly to MTF, resulting in the same number of installed IAB Nodes.

\item With intermediate minimum demands ($50-200$ Mb/s), PTF provides better peak throughputs by reducing the number of hops of the paths to TPs, this results into an increased number of installed IAB Nodes. This behavior does not emerge in MTF.

\item With small minimum demands ($0-50$ Mb/s), the number of installed devices is mainly driven by the objective of providing good coverage. Therefore, both formulations opt for a similar set of IAB Nodes. However, there is a difference on how these nodes are connected, as discussed in the next paragraphs.

\item As in the previous campaign, the lower part of the plot indicates that the impact of SRDs on throughput maximization is limited.

\end{enumerate*}

\begin{figure}[!ht]
\centering
\subfloat[DL rates]{\includegraphics[width=0.45\columnwidth]{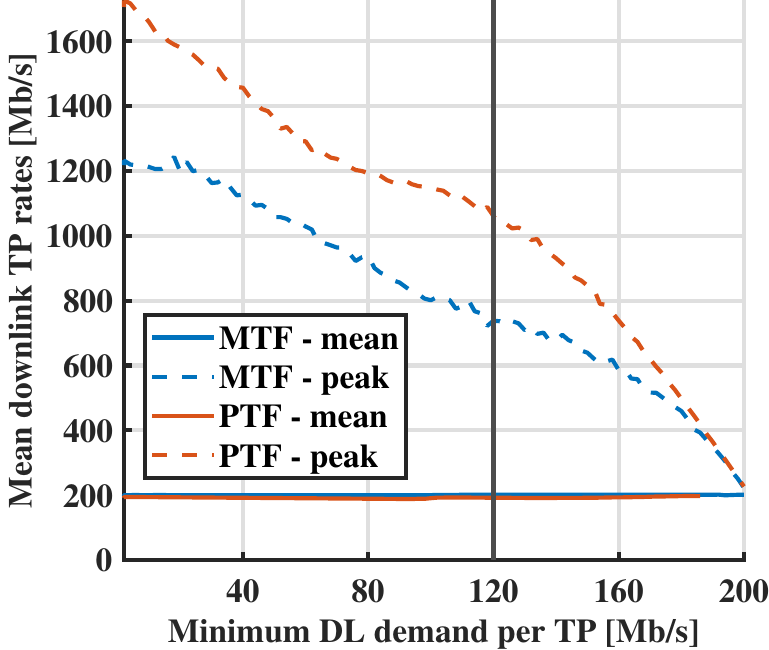}
\label{fig:dl_d}}
\hfil
\subfloat[UL rates]{\includegraphics[width=0.45\columnwidth]{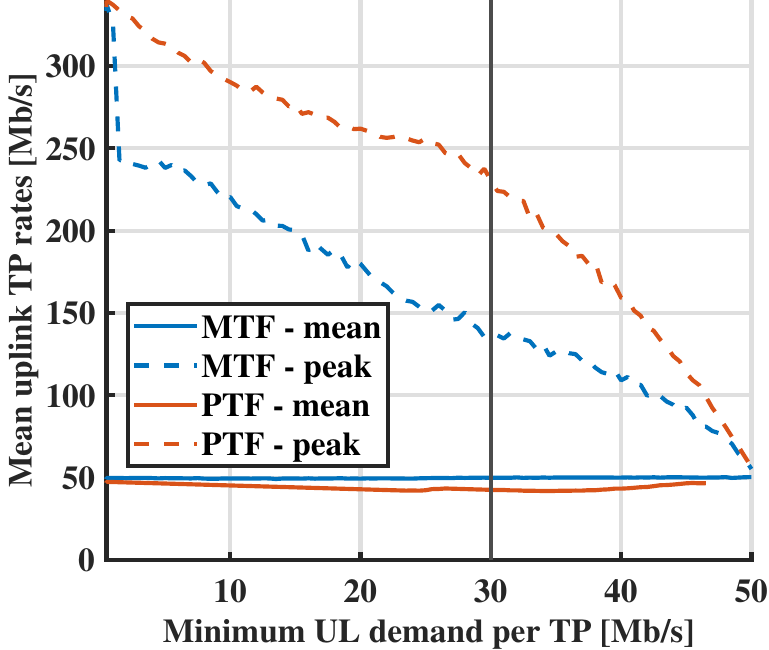}
\label{fig:ul_d}}
\caption{Overall throughput trends over guaranteed demand increase. \textcolor{black}{In both DL and UL transmissions, the peak throughput improvement of PTF over MTF degrades as the minimum guaranteed demand increases.}}
\label{fig:rates_d}
\end{figure}
Figure~\ref{fig:rates_d} presents the downlink (and uplink) per-TP rates considering, as in the previous campaign, peak-throughput and mean-throughput maximization in networks designed according to MTF and PTF. The offset between the peak throughputs achievable by the two formulations is consistent up to the two thirds of the minimum demand range. As resources are progressively consumed by minimum demands, PTF performance gradually aligns with the one of MTF, up to a threshold. At the value of $188$($47$) Mb/s for DL(UL), the heuristic approach used for PTF cannot find feasible solutions to accommodate the demands to be guaranteed. At the value of $200$($50$) Mb/s for DL(UL), even MTF cannot solve the planning problem, as the whole capacity available at the IAB Donor saturates and it can no longer guarantee the desired minimum demands.

\begin{figure}[!ht]
\centering
\subfloat[Number of hops from IAB Donor to UE]{\includegraphics[width=0.45\columnwidth]{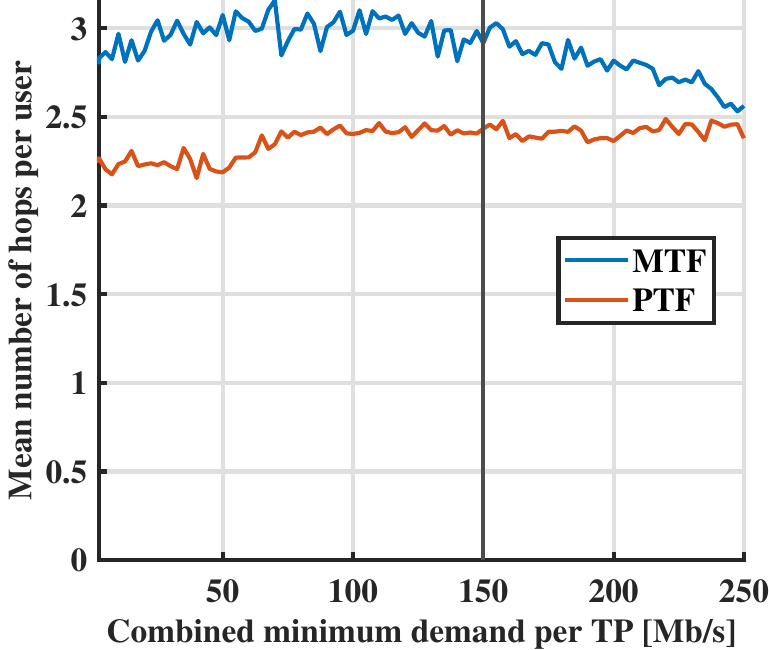}
\label{fig:mh_d}}
\hfil
\subfloat[Donor's degree]{\includegraphics[width=0.45\columnwidth]{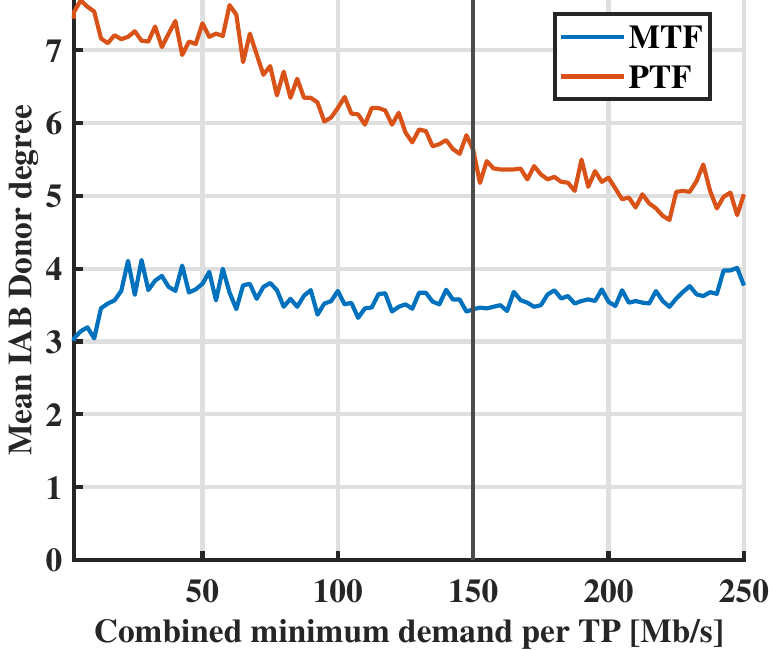}
\label{fig:dg_d}}
\caption{Variation of some topological properties over guaranteed demand increase. \textcolor{black}{A lower number of hops and higher Donor's degree can be observed for PTF compared to MTF. This difference slims down as the minimum demand increases, shrinking the solution space for the solver to explore.}}
\label{fig:topo_d}
\end{figure}

Figure~\ref{fig:mh_d} and~\ref{fig:dg_d} show the same topological properties as in Figure~\ref{fig:topo_b}. The reduction of the distance between MTF and PTF curves as the guaranteed minimum demand increases confirms the previous observations. However, we can better understand the difference between the two network layouts focusing on demands smaller than $50$ Mb/s. Although Figure~ \ref{fig:dev_d} shows the same number of installed devices, the PTF tends to interconnect them through shorter paths (in terms of hops) and preferring direct connections to the IAB Donor, thus further confirming that star-like topologies better suit to the peak-throughput maximization goal.

\begin{figure*}[!ht]
    \centering   \captionsetup{justification=centering,margin=1.5cm}
    \includegraphics[width=0.8\textwidth]{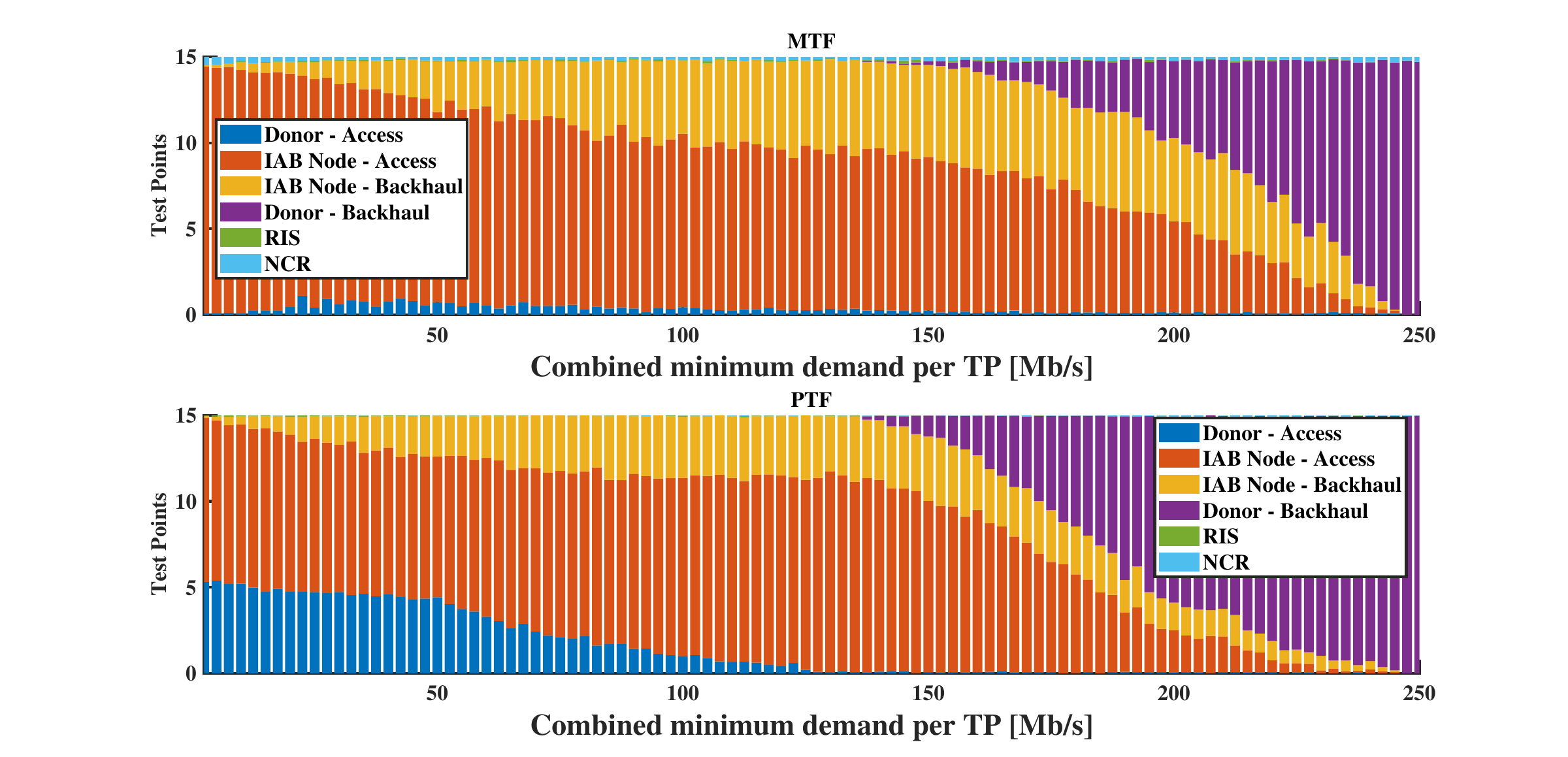}
    \caption{Distribution of bottlenecks over guaranteed demand variation. \textcolor{black}{Given the observed star-like topology trend, PTF has a larger concentration of Donor bottlenecks compared to MTF: as an access point at lower demands, and as a backhaul relay at higher ones.}}
    \label{fig:bn_d}
\end{figure*}
\begin{figure}[!hb]
\centering
\subfloat[MTF topology example, top view.]{\includegraphics[width=0.8\columnwidth]{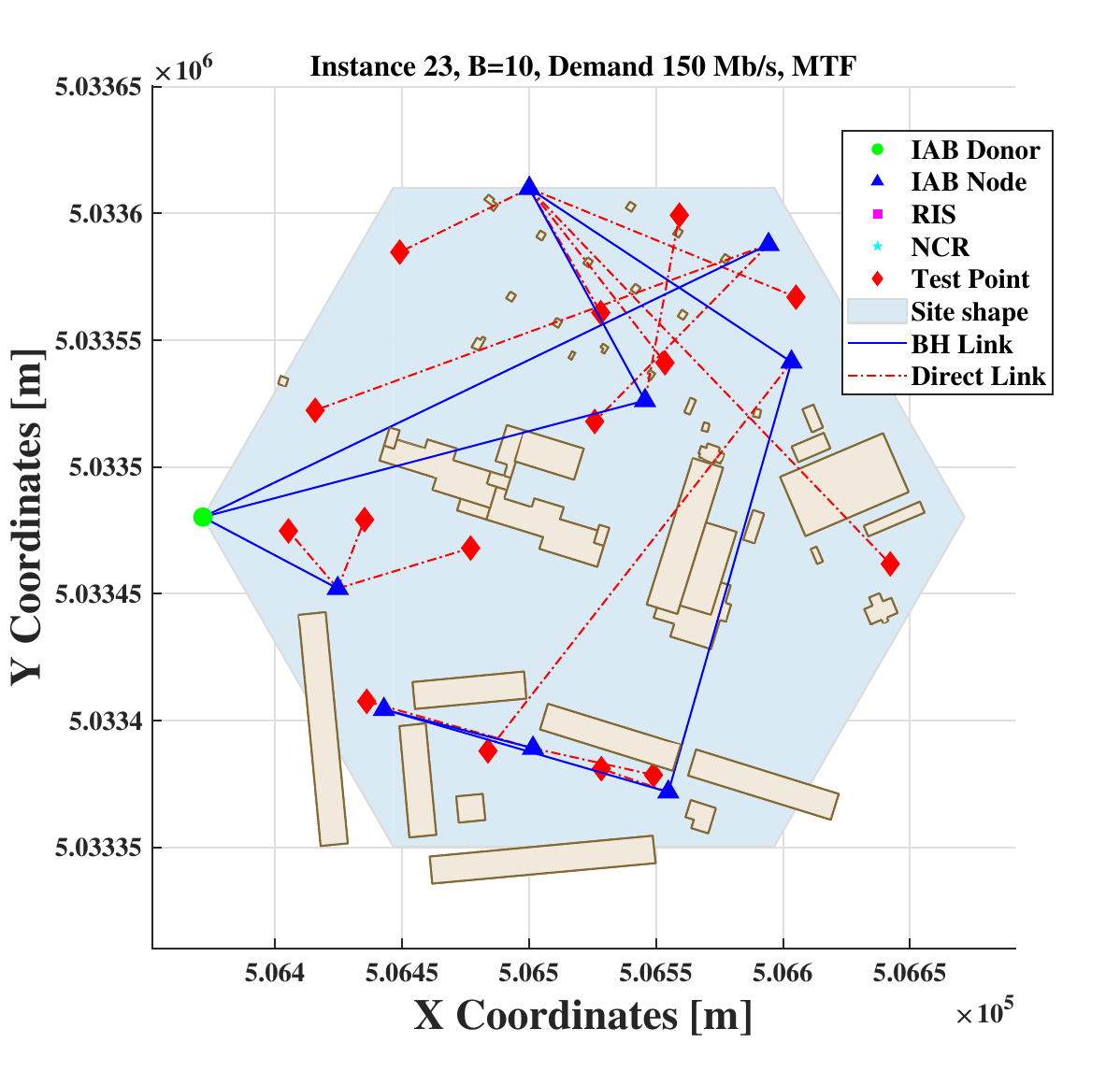}
\label{fig:MTF}}
\hfil
\subfloat[PTF topology example, top view.]{\includegraphics[width=0.8\columnwidth]{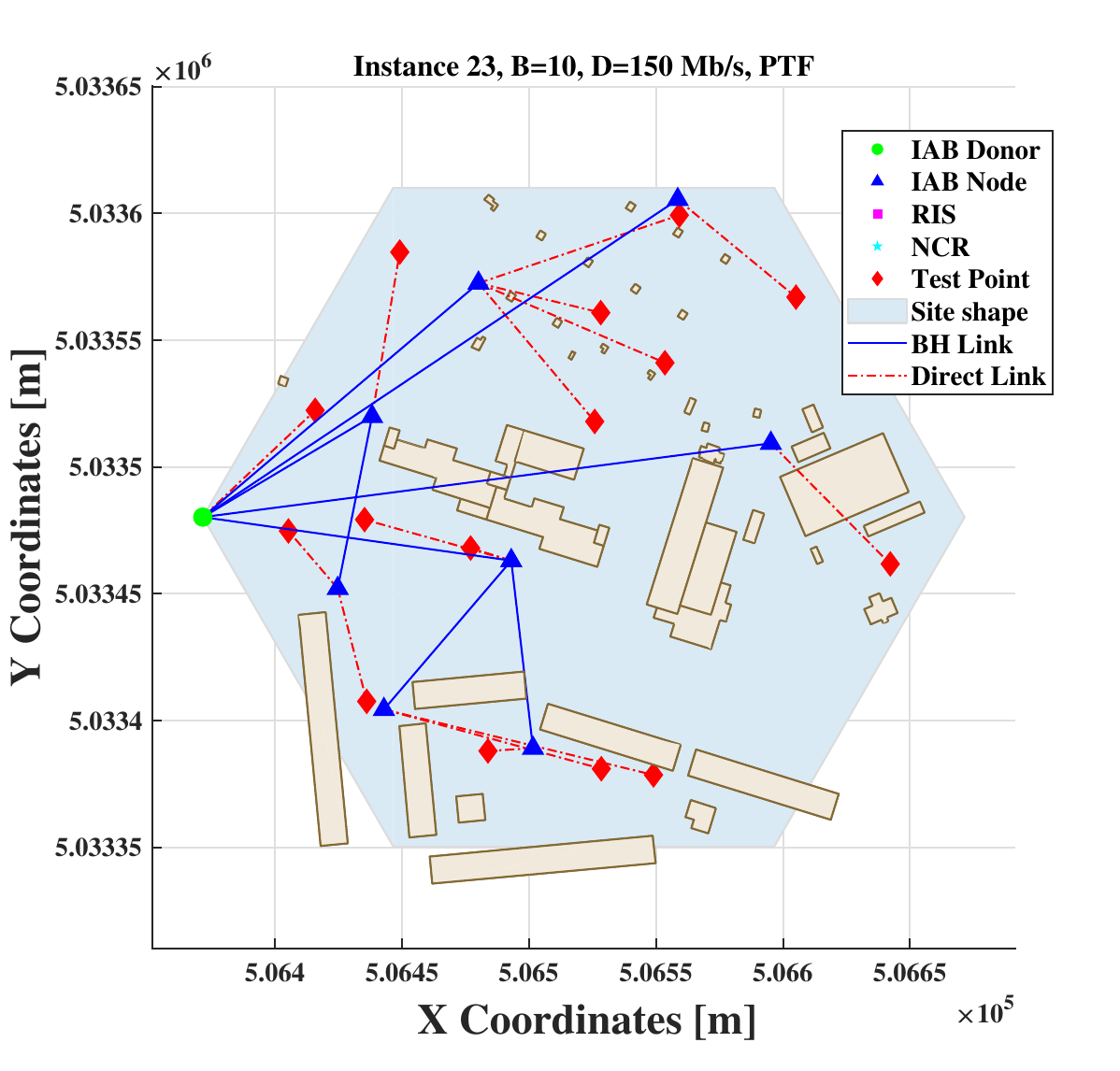}
\label{fig:PTF}}
\caption{Topology comparison of the two formulations from the top.}
\label{fig:comparison}
\vspace{-5mm}
\end{figure}
\begin{figure}[!tb]
\centering
\subfloat[MTF topology example, isometric view.]{\includegraphics[width=0.8\columnwidth]{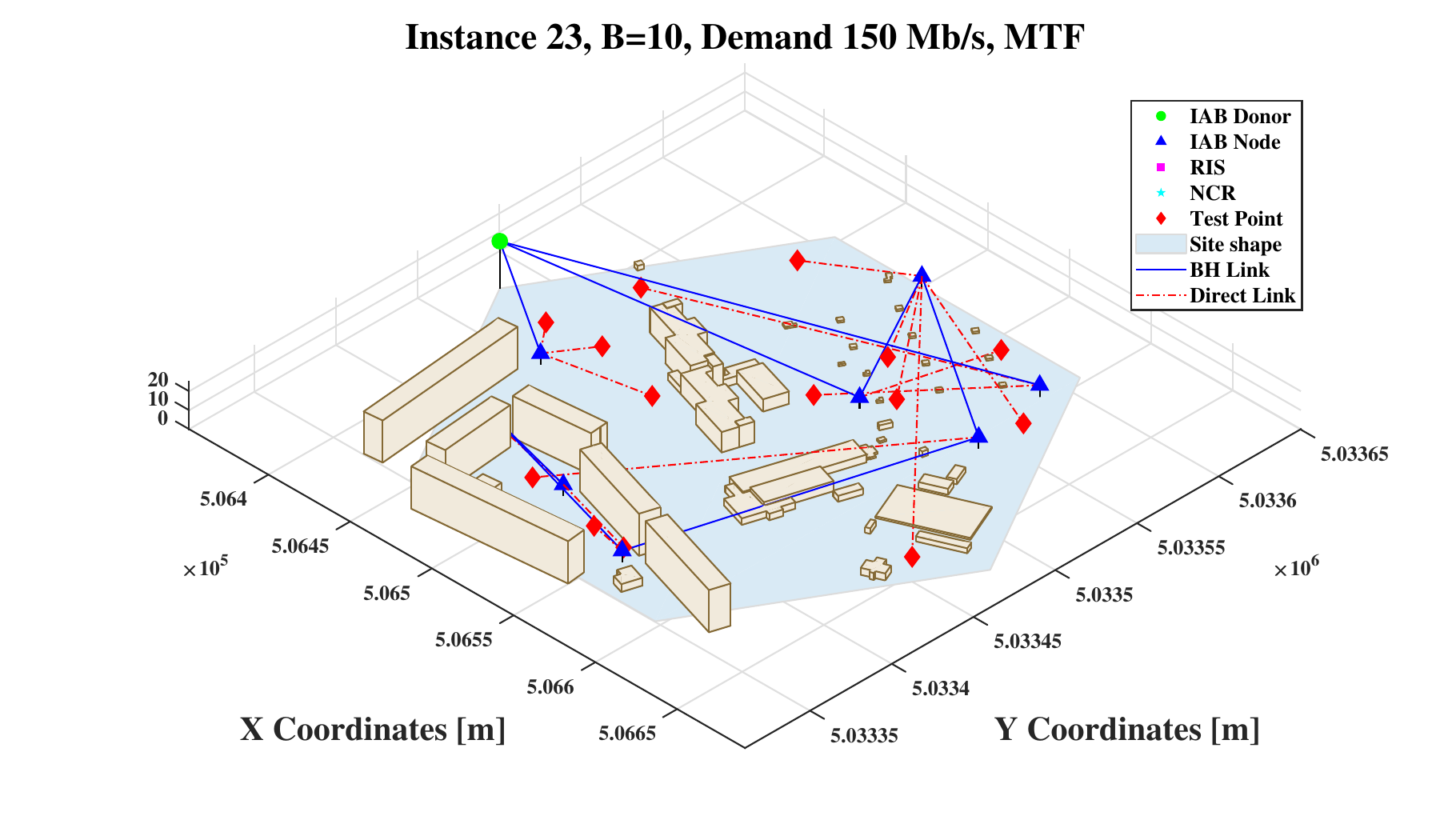}
\label{fig:MTF3d}}
\hfil
\subfloat[PTF topology example, isometric view.]{\includegraphics[width=0.8\columnwidth]{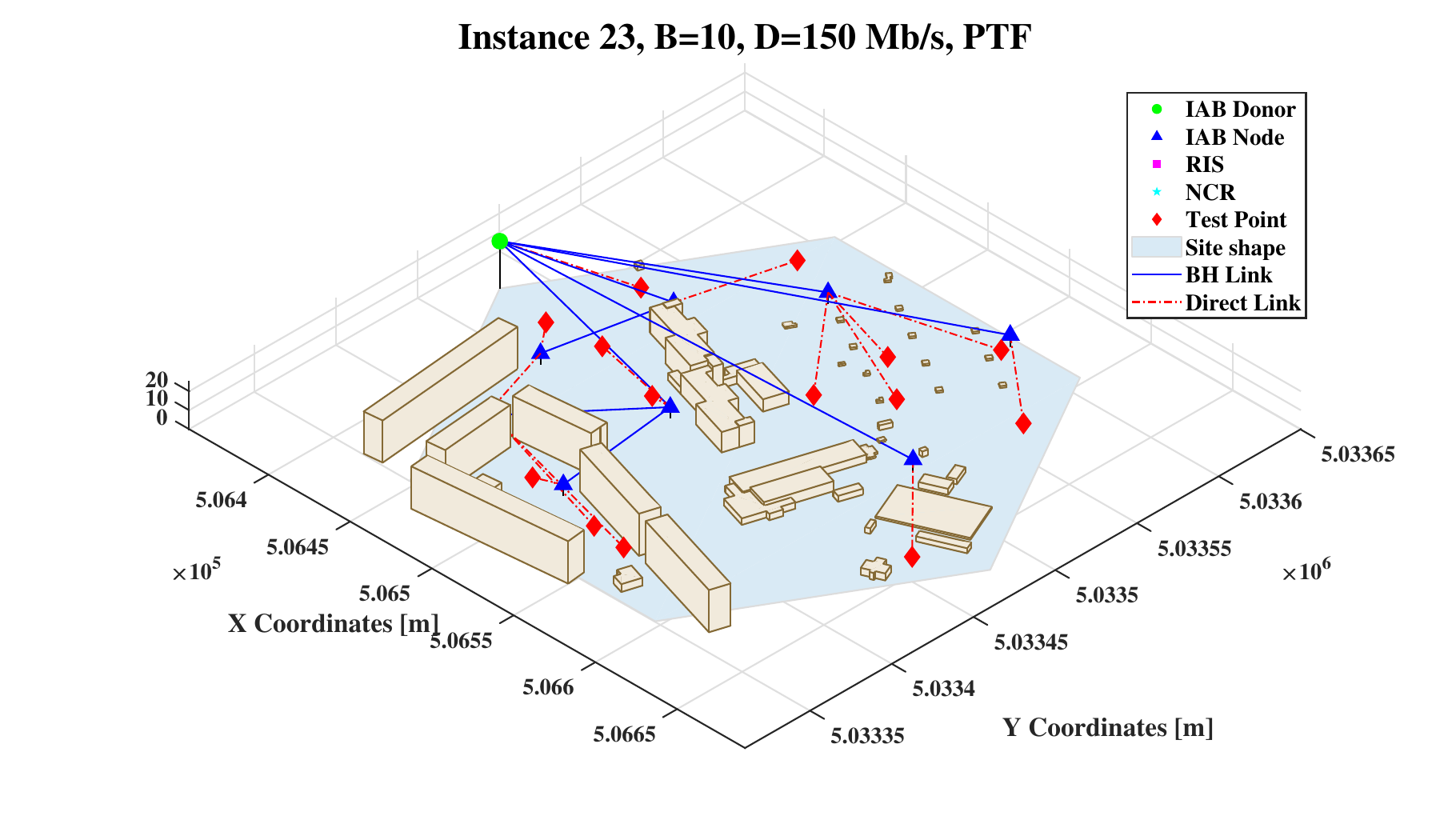}
\label{fig:PTF3d}}
\caption{Topology comparison of the two formulations from an isometric view.}
\label{fig:comparison3d}
\vspace{-5mm}
\end{figure}

In Figure~\ref{fig:bn_d}, the distribution of peak-traffic bottlenecks is analyzed. When the demand is low, PTF exhibits a fundamentally different behavior compared to MTF: oppositely to MTF, PTF connects as many as possible TPs directly to the IAB Donor. Consequently, the IAB Donor's access links frequently become the bottleneck of the network. As the demand increases, the behavior of the two approaches converges, as shown in the previous plots. However, when the guaranteed minimum demand increases approaching the saturation point, we can notice how PTF can shift bottlenecks up in the backhaul tree: moving them from IAB nodes' links to the IAB Donor's links.

\subsection{Visual comparison}
Figures~\ref{fig:comparison} and~\ref{fig:comparison3d} visually compare the layout differences of RANs generated by the two different formulations and deployed in the same cell area, from a top and an isometric view. The selected parameters are those of the intersection between the campaigns described in Sections~\ref{sub:budget} and~\ref{sub:demand}, with $10$ units of budget and $150$ Mb/s of overall guaranteed per-user demand. It is evident in Figures~\ref{fig:PTF}-\ref{fig:PTF3d} how PTF tends to install networks with fewer backhaul hops and larger IAB Donor's degree.

\section{Conclusion}
\label{sec:conclusion}
This study addresses the growing congestion in current mobile radio networks by investigating the potential of mmWave 5G networks in urban settings. It considers the way mmWave will change the interactions between users and networks, which, thanks to multi-Gbps wireless links, will be characterized by peak throughputs much larger than the average required per-user throughput. Therefore, the traffic will be mainly made of short bursts to transfer large volume of data and long idle periods to process them, as the dominance of video traffic in current mobile networks already shows.

The proposed MILP network planning models optimize operational parameters, emphasizing the maximization of achievable peak throughput and including the most recent mmWave paradigms of Integrated Access and Backhauling (IAB) and Smart Radio Environment (SRE). The results highlight the advantages of focusing on peak throughput during the network planning phase, providing insight into better accommodating the demands of mobile traffic without sacrificing the overall network capacity; \textcolor{black}{notably, peak throughput gains of up to 45\% in downlink and 70\% in uplink are achieved at equal budget, while maintaining comparable average throughput levels.}  This research contributes novel perspectives for the strategic deployment of mmWave 5G networks in urban scenarios, offering a basis for ongoing enhancements and refinements in mobile communication networks.

While the proposed approach operates at the strategic network planning level, complementary operational aspects such as dynamic scheduling and real-time resource allocation remain important areas for further investigation. Future research directions include extending the framework to encompass additional device types for more heterogeneous network environments, developing multi-objective optimization approaches that integrate sensing capabilities with communication objectives \cite{zhang2024target} for 6G's vision of Integrated Sensing and Communication (ISAC), and incorporating temporal dynamics such as mobility patterns and traffic variations into the planning optimization.

\section*{Acknowledgments}
The research in this paper has been carried out in the framework of
Huawei-Politecnico di Milano Joint Research Lab. The Authors
acknowledge Huawei Milan research center for the
collaboration.

The work of Ilario Filippini was supported by the EU under the Italian NRRP NextGenerationEU, partnership on “Telecommunications of the Future” (PE00000001 - program “RESTART”, Structural Project 6GWINET).

\bibliographystyle{IEEEtran}
\bibliography{bibliography.bib}

@article{jain2019impact,
  title={The impact of mobile blockers on millimeter wave cellular systems},
  author={Jain, Ish Kumar and Kumar, Rajeev and Panwar, Shivendra S},
  journal={IEEE Journal on Selected Areas in Communications},
  volume={37},
  number={4},
  pages={854--868},
  year={2019},
  publisher={IEEE}
}

@article{zhang2024target,
  title={Target detection and positioning aided by reconfigurable surfaces: Reflective or holographic?},
  author={Zhang, Xiaoyu and Zhang, Haobo and Liu, Liang and Han, Zhu and Poor, H Vincent and Di, Boya},
  journal={IEEE Transactions on Wireless Communications},
  year={2024},
  publisher={IEEE}
}

@article{gupta2022system,
  title={System-level analysis of full-duplex self-backhauled millimeter wave networks},
  author={Gupta, Manan and Roberts, Ian P and Andrews, Jeffrey G},
  journal={IEEE Transactions on Wireless Communications},
  volume={22},
  number={2},
  pages={1130--1144},
  year={2022},
  publisher={IEEE}
}

@ARTICLE{10649005,
  author={Kim, Jungjun and Lee, Jihwan and Bahk, Saewoong},
  journal={IEEE Transactions on Green Communications and Networking}, 
  title={Energy-Efficient and Robust Communication in Multi-Connectivity Enabled mmWave Ultra-Dense Networks}, 
  year={2024},
  volume={},
  number={},
  pages={1-1},
  doi={10.1109/TGCN.2024.3450524}}

@ARTICLE{7128403,
  author={García-Rois, Juan and Gómez-Cuba, Felipe and Akdeniz, Mustafa Riza and González-Castaño, Francisco J. and Burguillo, Juan C. and Rangan, Sundeep and Lorenzo, Beatriz},
  journal={IEEE Transactions on Wireless Communications}, 
  title={On the Analysis of Scheduling in Dynamic Duplex Multihop mmWave Cellular Systems}, 
  year={2015},
  volume={14},
  number={11},
  pages={6028-6042},
  doi={10.1109/TWC.2015.2446983}}

@ARTICLE{devoti2020,
  author={F. {Devoti} and I. {Filippini}},
  journal={IEEE/ACM Trans. on Networking}, 
  title={Planning mm-Wave Access Networks Under Obstacle Blockages: A Reliability-Aware Approach}, 
  year={2020},
  volume={28},
  number={5},
  pages={2203-2214},
  doi={10.1109/TNET.2020.3006926}
}

@article{Akdeniz2014,
archivePrefix = {arXiv},
arxivId = {1312.4921},
author = {Akdeniz, Mustafa Riza and Liu, Yuanpeng and Samimi, Mathew K. and Sun, Shu and Rangan, Sundeep and Rappaport, Theodore S. and Erkip, Elza},
doi = {10.1109/JSAC.2014.2328154},
eprint = {1312.4921},
issn = {07338716},
journal = {IEEE Jrn. on Sel. Areas in Comm.},
number = {6},
pages = {1164--1179},
publisher = {Institute of Electrical and Electronics Engineers Inc.},
title = {{Millimeter wave channel modeling and cellular capacity evaluation}},
volume = {32},
year = {2014}
}

@ARTICLE{rappaport2014,
  author={S. {Rangan} and T. S. {Rappaport} and E. {Erkip}},
  journal={Proceedings of the IEEE}, 
  title={Millimeter-Wave Cellular Wireless Networks: Potentials and Challenges}, 
  year={2014},
  volume={102},
  number={3},
  pages={366-385},
  doi={10.1109/JPROC.2014.2299397}
}

@INPROCEEDINGS{9771934,
  author={Fiore, Paolo and Moro, Eugenio and Filippini, Ilario and Capone, Antonio and De Donno, Danilo},
  booktitle={2022 IEEE Wireless Communications and Networking Conference (WCNC)}, 
  title={Boosting 5G mm-Wave IAB Reliability with Reconfigurable Intelligent Surfaces}, 
  year={2022},
  volume={},
  number={},
  pages={758-763},
  doi={10.1109/WCNC51071.2022.9771934}
}

@INPROCEEDINGS{Moro2021,
AUTHOR="E. Moro and I. Filippini and A. Capone and D. {De Donno}",
TITLE="Planning {Mm-Wave} Access Networks With Reconfigurable Intelligent Surfaces",
BOOKTITLE="2021 IEEE 32nd Annual Int. Sym. on Personal, Indoor and
Mobile Radio Comm. (PIMRC)",
ADDRESS="Helsinki, Finland",
DAYS=12,
MONTH=sep,
YEAR=2021,
}

@online{ericssonReportJun2025,
  author = {Ericsson},
  title = {"Ericsson Mobility Report: June 2025"},
  year = 2025,
  url = {https://www.ericsson.com/en/reports-and-papers/mobility-report/reports/june-2025},
  urldate = {}
}

@ARTICLE{10.3389/frcmn.2021.647284,  
AUTHOR={Zhang, Yongqiang and Kishk, Mustafa A. and Alouini, Mohamed-Slim},   	 
TITLE={A Survey on Integrated Access and Backhaul Networks},      	
JOURNAL={Frontiers in Communications and Networks},      	
VOLUME={2},           	
YEAR={2021},      	  
URL={https://www.frontiersin.org/articles/10.3389/frcmn.2021.647284},       	
DOI={10.3389/frcmn.2021.647284},      	
ISSN={2673-530X}
}

@INPROCEEDINGS{9930587,
  author={Leone, Giuseppe and Moro, Eugenio and Filippini, Ilario and Capone, Antonio and Donno, Danilo De},
  booktitle={2022 20th International Symposium on Modeling and Optimization in Mobile, Ad hoc, and Wireless Networks (WiOpt)}, 
  title={Towards Reliable mmWave 6G RAN: Reconfigurable Surfaces, Smart Repeaters, or Both?}, 
  year={2022},
  volume={},
  number={},
  pages={81-88},
  doi={10.23919/WiOpt56218.2022.9930587}
}

@online{sandvineReport2024,
  author = {Sandvine},
  title = {"2024 Global Internet Phenomena Report"},
  year = 2024,
  url = {https://www.sandvine.com/global-internet-phenomena-report-2024},
  urldate = {}
}

@techreport{3GPPTR38.874,
  title={3rd Generation Partnership Project;
Technical Specification Group Radio Access Network;
NR;
Study on Integrated Access and Backhaul;
(Release 16)},
  author={3GPP},
  institution={3GPP},
  journal={Technical Report TR 38.874
V16.0.0},
  publisher={3GPP},
  year={2018}
}

@techreport{GRANW2018study,
  title={Study on Channel Model for Frequencies From
0.5 to 100 GHz (Release 15)},
  author={G. R. A. N. W. Group},
  institution={3GPP},
  journal={document TR 38.901},
  publisher={3GPP},
  year={2020}
}

@online{MilanoGeoportale,
  organization = {Comune di Milano},
  title = {"Milano Geoportale - Open Data"},
  year = 2012,
  url = {https://geoportale.comune.milano.it/sit/open-data/},
  urldate = {}
}

@INPROCEEDINGS{8254900,
  author={MacCartney, George R. and Rappaport, Theodore S. and Rangan, Sundeep},
  booktitle={GLOBECOM 2017 - 2017 IEEE Global Communications Conference}, 
  title={Rapid Fading Due to Human Blockage in Pedestrian Crowds at 5G Millimeter-Wave Frequencies}, 
  year={2017},
  volume={},
  number={},
  pages={1-7},
  doi={10.1109/GLOCOM.2017.8254900}
}

@INPROCEEDINGS{8514996,
  author={Polese, Michele and Giordani, Marco and Roy, Arnab and Goyal, Sanjay and Castor, Douglas and Zorzi, Michele},
  booktitle={2018 IEEE 23rd International Workshop on Computer Aided Modeling and Design of Communication Links and Networks (CAMAD)}, 
  title={End-to-End Simulation of Integrated Access and Backhaul at mmWaves}, 
  year={2018},
  volume={},
  number={},
  pages={1-7},
  doi={10.1109/CAMAD.2018.8514996}
}

@ARTICLE{9040265,
  author={Polese, Michele and Giordani, Marco and Zugno, Tommaso and Roy, Arnab and Goyal, Sanjay and Castor, Douglas and Zorzi, Michele},
  journal={IEEE Communications Magazine}, 
  title={Integrated Access and Backhaul in 5G mmWave Networks: Potential and Challenges}, 
  year={2020},
  volume={58},
  number={3},
  pages={62-68},
  doi={10.1109/MCOM.001.1900346}
}

@article{tsilipakos2020toward,
  title={Toward Intelligent Metasurfaces: The Progress from Globally Tunable Metasurfaces to Software-Defined Metasurfaces with an Embedded Network of Controllers},
  author={Tsilipakos, Odysseas and Tasolamprou, Anna C and Pitilakis, Alexandros and Liu, Fu and Wang, Xuchen and Mirmoosa, Mohammad Sajjad and Tzarouchis, Dimitrios C and Abadal, Sergi and Taghvaee, Hamidreza and Liaskos, Christos and others},
  journal={Advanced Optical Materials},
  volume={8},
  number={17},
  pages={2000783},
  year={2020},
  publisher={Wiley Online Library}
}

@article{di2019smart,
  title={Smart radio environments empowered by reconfigurable AI meta-surfaces: An idea whose time has come},
  author={Di Renzo, Marco and Debbah, Merouane and Phan-Huy, Dinh-Thuy and Zappone, Alessio and Alouini, Mohamed-Slim and Yuen, Chau and Sciancalepore, Vincenzo and Alexandropoulos, George C and Hoydis, Jakob and Gacanin, Haris and others},
  journal={EURASIP Journal on Wireless Communications and Networking},
  volume={2019},
  number={1},
  pages={1--20},
  year={2019},
  publisher={SpringerOpen}
}

@ARTICLE{6678102,
  author={Ju, Hyungsik and Zhang, Rui},
  journal={IEEE Transactions on Wireless Communications}, 
  title={Throughput Maximization in Wireless Powered Communication Networks}, 
  year={2014},
  volume={13},
  number={1},
  pages={418-428},
  doi={10.1109/TWC.2013.112513.130760}
}

@ARTICLE{7342886,
  author={Kutty, Shajahan and Sen, Debarati},
  journal={IEEE Communications Surveys \& Tutorials}, 
  title={Beamforming for Millimeter Wave Communications: An Inclusive Survey}, 
  year={2016},
  volume={18},
  number={2},
  pages={949-973},
  doi={10.1109/COMST.2015.2504600}
}

@ARTICLE{9720231,
  author={Flamini, Roberto and De Donno, Danilo and Gambini, Jonathan and Giuppi, Francesco and Mazzucco, Christian and Milani, Angelo and Resteghini, Laura},
  journal={IEEE Transactions on Antennas and Propagation}, 
  title={Toward a Heterogeneous Smart Electromagnetic Environment for Millimeter-Wave Communications: An Industrial Viewpoint}, 
  year={2022},
  volume={70},
  number={10},
  pages={8898-8910},
  doi={10.1109/TAP.2022.3151978}
}

@article{niu2015survey,
  title={A survey of millimeter wave communications (mmWave) for 5G: opportunities and challenges},
  author={Niu, Yong and Li, Yong and Jin, Depeng and Su, Li and Vasilakos, Athanasios V},
  journal={Wireless networks},
  volume={21},
  pages={2657--2676},
  year={2015},
  publisher={Springer}
}

@ARTICLE{8700132,
  author={Lopez, Aida Vera and Chervyakov, Andrey and Chance, Greg and Verma, Sumit and Tang, Yang},
  journal={IEEE Wireless Communications}, 
  title={Opportunities and Challenges of mmWave NR}, 
  year={2019},
  volume={26},
  number={2},
  pages={4-6},
  doi={10.1109/MWC.2019.8700132}
}

@ARTICLE{8373698,
  author={Wang, Xiong and Kong, Linghe and Kong, Fanxin and Qiu, Fudong and Xia, Mingyu and Arnon, Shlomi and Chen, Guihai},
  journal={IEEE Communications Surveys \& Tutorials}, 
  title={Millimeter Wave Communication: A Comprehensive Survey}, 
  year={2018},
  volume={20},
  number={3},
  pages={1616-1653},
  doi={10.1109/COMST.2018.2844322}
}

@article{sakaguchi2017and,
  title={Where, when, and how mmWave is used in 5G and beyond},
  author={Sakaguchi, Kei and Haustein, Thomas and Barbarossa, Sergio and Strinati, Emilio Calvanese and Clemente, Antonio and Destino, Giuseppe and P{\"a}rssinen, Aarno and Kim, Ilgyu and Chung, Heesang and Kim, Junhyeong and others},
  journal={IEICE Transactions on Electronics},
  volume={100},
  number={10},
  pages={790--808},
  year={2017},
  publisher={The Institute of Electronics, Information and Communication Engineers}
}

@ARTICLE{7959169,
  author={Xiao, Ming and Mumtaz, Shahid and Huang, Yongming and Dai, Linglong and Li, Yonghui and Matthaiou, Michail and Karagiannidis, George K. and Björnson, Emil and Yang, Kai and I, Chih-Lin and Ghosh, Amitabha},
  journal={IEEE Journal on Selected Areas in Communications}, 
  title={Millimeter Wave Communications for Future Mobile Networks}, 
  year={2017},
  volume={35},
  number={9},
  pages={1909-1935},
  doi={10.1109/JSAC.2017.2719924}
}

@ARTICLE{9187867,
  author={Madapatha, Charitha and Makki, Behrooz and Fang, Chao and Teyeb, Oumer and Dahlman, Erik and Alouini, Mohamed-Slim and Svensson, Tommy},
  journal={IEEE Open Journal of the Communications Society}, 
  title={On Integrated Access and Backhaul Networks: Current Status and Potentials}, 
  year={2020},
  volume={1},
  number={},
  pages={1374-1389},
  doi={10.1109/OJCOMS.2020.3022529}
}

@ARTICLE{8882288,
  author={Saha, Chiranjib and Dhillon, Harpreet S.},
  journal={IEEE Journal on Selected Areas in Communications}, 
  title={Millimeter Wave Integrated Access and Backhaul in 5G: Performance Analysis and Design Insights}, 
  year={2019},
  volume={37},
  number={12},
  pages={2669-2684},
  doi={10.1109/JSAC.2019.2947997}
}

@ARTICLE{9433519,
  author={Cudak, Mark and Ghosh, Amitabha and Ghosh, Arunabha and Andrews, Jeffrey},
  journal={IEEE Communications Magazine}, 
  title={Integrated Access and Backhaul: A Key Enabler for 5G Millimeter-Wave Deployments}, 
  year={2021},
  volume={59},
  number={4},
  pages={88-94},
  doi={10.1109/MCOM.001.2000690}
}

@ARTICLE{9086766,
  author={ElMossallamy, Mohamed A. and Zhang, Hongliang and Song, Lingyang and Seddik, Karim G. and Han, Zhu and Li, Geoffrey Ye},
  journal={IEEE Transactions on Cognitive Communications and Networking}, 
  title={Reconfigurable Intelligent Surfaces for Wireless Communications: Principles, Challenges, and Opportunities}, 
  year={2020},
  volume={6},
  number={3},
  pages={990-1002},
  doi={10.1109/TCCN.2020.2992604}
}

@ARTICLE{8796365,
  author={Basar, Ertugrul and Di Renzo, Marco and De Rosny, Julien and Debbah, Merouane and Alouini, Mohamed-Slim and Zhang, Rui},
  journal={IEEE Access}, 
  title={Wireless Communications Through Reconfigurable Intelligent Surfaces}, 
  year={2019},
  volume={7},
  number={},
  pages={116753-116773},
  doi={10.1109/ACCESS.2019.2935192}
}

@ARTICLE{9119122,
  author={Di Renzo, Marco and Ntontin, Konstantinos and Song, Jian and Danufane, Fadil H. and Qian, Xuewen and Lazarakis, Fotis and De Rosny, Julien and Phan-Huy, Dinh-Thuy and Simeone, Osvaldo and Zhang, Rui and Debbah, Meroaune and Lerosey, Geoffroy and Fink, Mathias and Tretyakov, Sergei and Shamai, Shlomo},
  journal={IEEE Open Journal of the Communications Society}, 
  title={Reconfigurable Intelligent Surfaces vs. Relaying: Differences, Similarities, and Performance Comparison}, 
  year={2020},
  volume={1},
  number={},
  pages={798-807},
  doi={10.1109/OJCOMS.2020.3002955}
}

@ARTICLE{9122596,
  author={Gong, Shimin and Lu, Xiao and Hoang, Dinh Thai and Niyato, Dusit and Shu, Lei and Kim, Dong In and Liang, Ying-Chang},
  journal={IEEE Communications Surveys \& Tutorials}, 
  title={Toward Smart Wireless Communications via Intelligent Reflecting Surfaces: A Contemporary Survey}, 
  year={2020},
  volume={22},
  number={4},
  pages={2283-2314},
  doi={10.1109/COMST.2020.3004197}
}

@INPROCEEDINGS{10065985,
  author={Xin, Jincan and Xu, Sen and Xiong, Shangkun and Xu, Hua and Zhang, Hua},
  booktitle={2022 IEEE 8th International Conference on Computer and Communications (ICCC)}, 
  title={A Survey on Network Controlled Repeater Technology}, 
  year={2022},
  volume={},
  number={},
  pages={1097-1101},
  doi={10.1109/ICCC56324.2022.10065985}
}

@article{da2023system,
  title={System Level Evaluation of Network-Controlled Repeaters: Performance Improvement of Serving Cell and Interference Impact on Neighbor Cells},
  author={da Silva, Gabriel and Falc{\~a}o, Erik RB and Monteiro, Victor F and Moreira, Darlan C and Sousa, Diego A and Maciel, Tarcisio F and Lima, Fco and Rafael, M and Makki, Behrooz},
  journal={XLI Brazilian Symposium on Telecommunications and Signal Processing (SBrT)},
  year={2023}
}

@inproceedings{rao2011network,
  title={Network characteristics of video streaming traffic},
  author={Rao, Ashwin and Legout, Arnaud and Lim, Yeon-sup and Towsley, Don and Barakat, Chadi and Dabbous, Walid},
  booktitle={Proceedings of the seventh conference on emerging networking experiments and technologies},
  pages={1--12},
  year={2011}
}

@INPROCEEDINGS{10168854,
  author={Ayoubi, Reza Aghazadeh and Mizmizi, Marouan and Tagliaferri, Dario and Donno, Danilo De and Spagnolini, Umberto},
  booktitle={2023 21st Mediterranean Communication and Computer Networking Conference (MedComNet)}, 
  title={Network-Controlled Repeaters vs. Reconfigurable Intelligent Surfaces for 6G mmW Coverage Extension: A Simulative Comparison}, 
  year={2023},
  volume={},
  number={},
  pages={196-202},
  doi={10.1109/MedComNet58619.2023.10168854}
}

@article{raghavan2019statistical,
  title={Statistical blockage modeling and robustness of beamforming in millimeter-wave systems},
  author={Raghavan, Vasanthan and Akhoondzadeh-Asl, Lida and Podshivalov, Vladimir and Hulten, Joakim and Tassoudji, Mohammad Ali and Koymen, Ozge Hizir and Sampath, Ashwin and Li, Junyi},
  journal={IEEE Transactions on Microwave Theory and Techniques},
  volume={67},
  number={7},
  pages={3010--3024},
  year={2019},
  publisher={IEEE}
}

@ARTICLE{6515173,
  author={Rappaport, Theodore S. and Sun, Shu and Mayzus, Rimma and Zhao, Hang and Azar, Yaniv and Wang, Kevin and Wong, George N. and Schulz, Jocelyn K. and Samimi, Mathew and Gutierrez, Felix},
  journal={IEEE Access}, 
  title={Millimeter Wave Mobile Communications for 5G Cellular: It Will Work!}, 
  year={2013},
  volume={1},
  number={},
  pages={335-349},
  doi={10.1109/ACCESS.2013.2260813}
}

@INPROCEEDINGS{10293769,
  author={Fiore, Paolo and Filippini, Ilario and De Donno, Danilo},
  booktitle={2023 IEEE 34th Annual International Symposium on Personal, Indoor and Mobile Radio Communications (PIMRC)}, 
  title={Shaping Next-Generation RAN Topologies to Meet Future Traffic Demands: A Peak Throughput Study}, 
  year={2023},
  volume={},
  number={},
  pages={1-7},
  doi={10.1109/PIMRC56721.2023.10293769}
}

@INPROCEEDINGS{6290250,
  author={Bai, Tianyang and Vaze, Rahul and Heath, Robert W.},
  booktitle={2012 International Conference on Signal Processing and Communications (SPCOM)}, 
  title={Using random shape theory to model blockage in random cellular networks}, 
  year={2012},
  volume={},
  number={},
  pages={1-5},
  doi={10.1109/SPCOM.2012.6290250}
}

@techreport{3GPPTR38.867,
  title={3rd Generation Partnership Project;
Technical Specification Group Radio Access Network;
NR;
Study on NR Network-controlled Repeaters;
(Release 18)},
  author={3GPP},
  institution={3GPP},
  journal={Technical Report TR 38.867
V18.0.0},
  publisher={3GPP},
  year={2022}
}

@techreport{ETSIGRRIS001,
  title={Reconfigurable Intelligent Surfaces (RIS);
Use Cases, Deployment Scenarios and Requirements},
  author={ETSI},
  institution={ETSI},
  journal={Group Report RIS 001 V1.1.1},
  publisher={ETSI},
  year={2023}
}

@techreport{ETSIGRRIS002,
  title={Reconfigurable Intelligent Surfaces (RIS);
         Technological challenges, architecture and
impact on standardization},
  author={ETSI},
  institution={ETSI},
  journal={Group Report RIS 002 V1.1.1},
  publisher={ETSI},
  year={2023}
}

@techreport{ETSIGRRIS003,
  title={Reconfigurable Intelligent Surfaces (RIS);
Communication Models, Channel Models, Channel Estimation and Evaluation Methodology},
  author={ETSI},
  institution={ETSI},
  journal={Group Report RIS 003 V1.1.1},
  publisher={ETSI},
  year={2023}
}

@ARTICLE{6840343,
  author={Bai, Tianyang and Vaze, Rahul and Heath, Robert W.},
  journal={IEEE Transactions on Wireless Communications}, 
  title={Analysis of Blockage Effects on Urban Cellular Networks}, 
  year={2014},
  volume={13},
  number={9},
  pages={5070-5083},
  doi={10.1109/TWC.2014.2331971}
}

@inproceedings{alwajeeh2017high,
  title={A high-speed 2.5 D ray-tracing propagation model for microcellular systems, application: Smart cities},
  author={Alwajeeh, Taha and Combeau, Pierre and Vauzelle, Rodolphe and Bounceur, Ahcene},
  booktitle={2017 11th European Conference on Antennas and Propagation (EUCAP)},
  pages={3515--3519},
  year={2017},
  organization={IEEE}
}

@article{liu2013full,
  title={Full automatic preprocessing of digital map for 2.5 D ray tracing propagation model in urban microcellular environment},
  author={Liu, Zhong-Yu and Guo, Li-Xin and Tao, Wei},
  journal={Waves in Random and Complex Media},
  volume={23},
  number={3},
  pages={267--278},
  year={2013},
  publisher={Taylor \& Francis}
}

@inproceedings{maccartney2016millimeter,
  title={Millimeter-wave human blockage at 73 GHz with a simple double knife-edge diffraction model and extension for directional antennas},
  author={MacCartney, George R and Deng, Sijia and Sun, Shu and Rappaport, Theodore S},
  booktitle={2016 IEEE 84th Vehicular Technology Conference (VTC-Fall)},
  pages={1--6},
  year={2016},
  organization={IEEE}
}

@article{collonge2004influence,
  title={Influence of the human activity on wide-band characteristics of the 60 GHz indoor radio channel},
  author={Collonge, Sylvain and Zaharia, Gheorghe and Zein, GE},
  journal={IEEE Transactions on Wireless Communications},
  volume={3},
  number={6},
  pages={2396--2406},
  year={2004},
  publisher={IEEE}
}

@article{zhang2021design,
  title={Design of full duplex millimeter-wave integrated access and backhaul networks},
  author={Zhang, Junkai and Garg, Navneet and Holm, Mark and Ratnarajah, Tharmalingam},
  journal={IEEE Wireless Communications},
  volume={28},
  number={1},
  pages={60--67},
  year={2021},
  publisher={IEEE}
}

@inproceedings{albanese2022ris,
  title={RIS-aware indoor network planning: The Rennes railway station case},
  author={Albanese, Antonio and Encinas-Lago, Guillermo and Sciancalepore, Vincenzo and Costa-P{\'e}rez, Xavier and Phan-Huy, Dinh-Thuy and Ros, St{\'e}phane},
  booktitle={ICC 2022-IEEE International Conference on Communications},
  pages={2028--2034},
  year={2022},
  organization={IEEE}
}

@article{zhang2024mobility,
  title={Mobility-aware resource allocation for mmWave IAB networks: A multi-agent reinforcement learning approach},
  author={Zhang, Bibo and Filippini, Ilario},
  journal={IEEE/ACM Transactions on Networking},
  volume={32},
  number={4},
  pages={3559--3574},
  year={2024},
  publisher={IEEE}
}

@article{santana20245g,
  title={5G mmWave Network Planning Using Machine Learning for Path Loss Estimation},
  author={Santana, Yosvany Hervis and Alonso, Rodney Martinez and Nieto, Glauco Guillen and Martens, Luc and Joseph, Wout and Plets, David},
  journal={IEEE Open Journal of the Communications Society},
  volume={5},
  pages={3451--3467},
  year={2024},
  publisher={IEEE}
}

@article{ayoubi2025optimal,
  title={Optimal planning for heterogeneous smart radio environments},
  author={Ayoubi, Reza Agahzadeh and Moro, Eugenio and Mizmizi, Marouan and Tagliaferri, Dario and Filippini, Ilario and Spagnolini, Umberto},
  journal={IEEE Transactions on Mobile Computing},
  year={2025},
  publisher={IEEE}
}

@article{dai2024user,
  title={User association and channel allocation in 5G mobile asymmetric Multi-Band heterogeneous networks},
  author={Dai, Miao and Sun, Gang and Yu, Hongfang and Wang, Sheng and Niyato, Dusit},
  journal={IEEE Transactions on Mobile Computing},
  year={2024},
  publisher={IEEE}
}


\section{Biography Section}
 
\vspace{-50pt}
\begin{IEEEbiography}[{\includegraphics[width=1in,height=1.25in,clip,keepaspectratio]{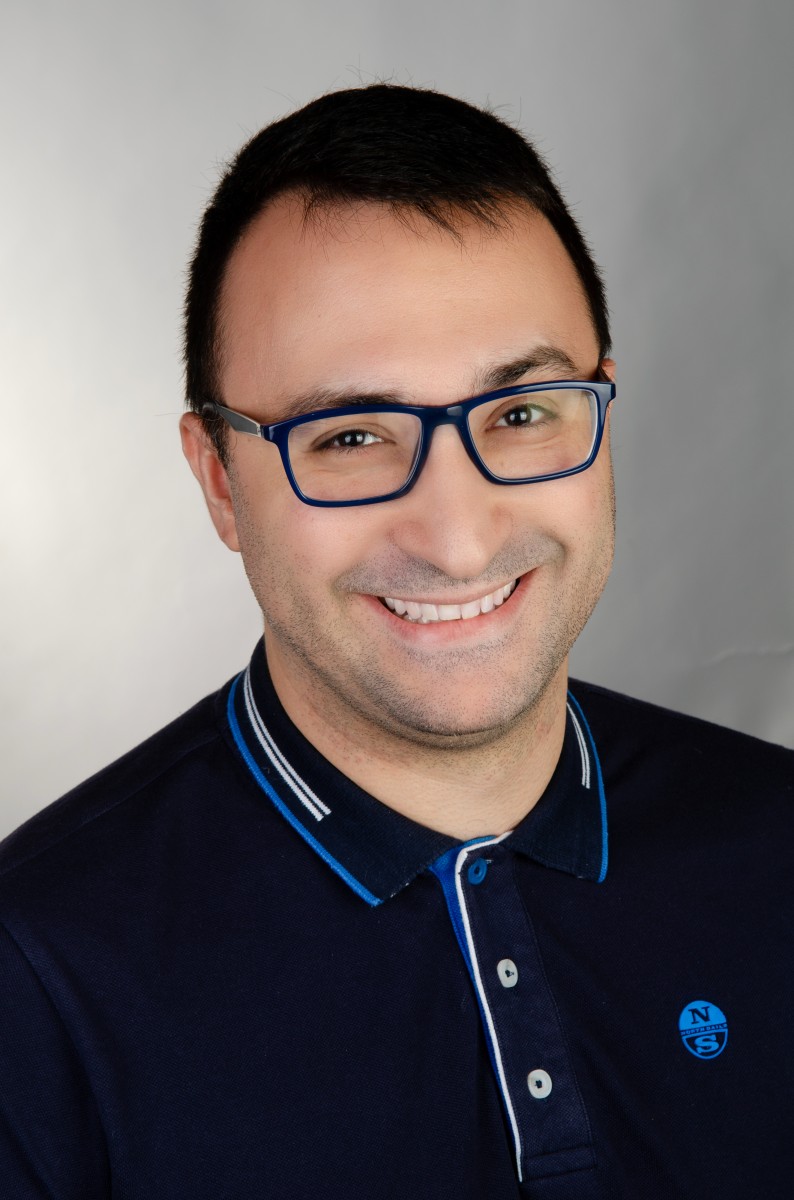}}]{Paolo Fiore}
(Student Member, IEEE) received his BS in computer science and engineering in 2013 and his MS in telecommunications engineering in 2021, both from Politecnico di Milano. He is currently a visiting researcher at NEC Laboratories Europe, in Heidelberg, Germany. 
His research interests consist in resource management, optimization and localization for wireless networks in reconfigurable environments, encompassing network planning and smart radio devices. He was the corecipient of the Best Paper Award at IEEE PIMRC 2023.
\end{IEEEbiography}

\vspace{-50pt}

\begin{IEEEbiography}[{\includegraphics[width=1in,height=1.25in,clip,keepaspectratio]{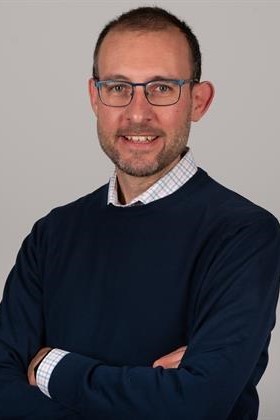}}]{Ilario Filippini}
(Senior Member, IEEE) is an Associate Professor at the Dipartimento di Elettronica, Informazione e Bioingegneria of Politecnico di Milano. He received an M.S. in Telecommunication Engineering and a Ph.D. in Information Engineering from Politecnico di Milano in 2005 and 2009, respectively. From February 2008 to August 2008, he was visiting the Department of Electrical and Computer Engineering of The Ohio State University in Columbus (OH) working on routing in Cognitive Radio Networks.
He works on networking topics, his main research activities include radio resource management and optimization in wireless networks, programmable networks, and smart radio environments. He has co-authored more than 70 peer-reviewed conference and journal papers. He serves in the Technical Program Committee of major conferences in Networking and as an Editor of Computer Networks (Elsevier).
\end{IEEEbiography}

\vspace{-50pt}

\begin{IEEEbiography}[{\includegraphics[width=1in,height=1.25in,clip,keepaspectratio]{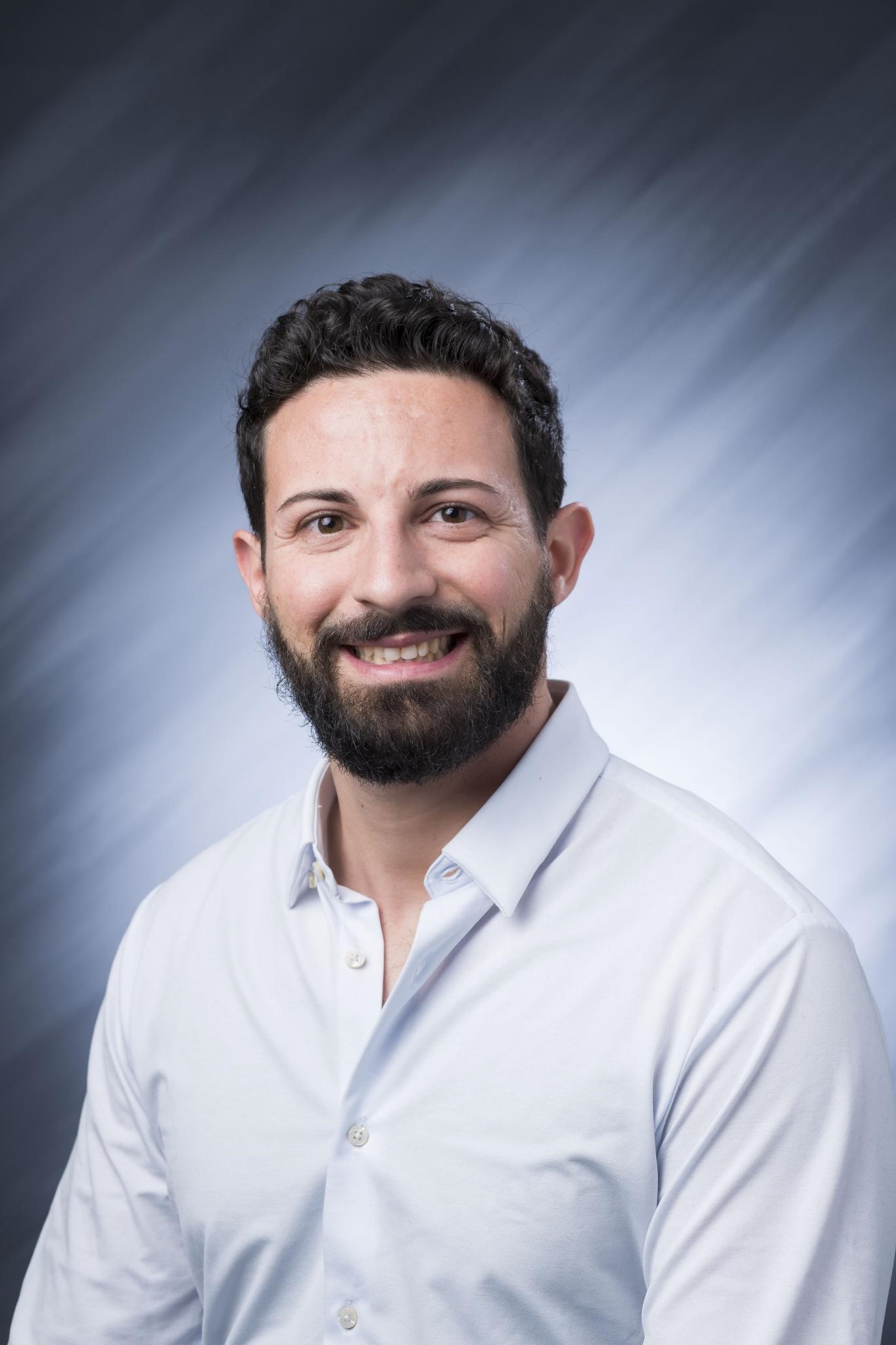}}]{Danilo De Donno}
received the B.Sc. and M.Sc. degrees in telecommunication engineering from the Politecnico di Milano, Milan, Italy, in 2005 and 2008, respectively, and the Ph.D. degree in information engineering from the University of Salento, Lecce, Italy, in 2012. He was a Post-Doctoral Fellow with the Electromagnetic Lab Lecce (EML2), University of Salento, from 2012 to 2015, and a Post-Doctoral Researcher with the IMDEA Networks Institute, Madrid, Spain, from 2015 to 2017. In July 2017, he joined the Huawei Research Center, Milan, as a Wireless System Engineer. His areas of interests include mmWave communications, with main research, focus on network-level analysis of 5G mmWave systems, integrated access and backhaul (IAB), PHY, and MAC algorithms for hybrid and full-digital system architectures.
\end{IEEEbiography}

\newpage
{\appendix[Modified 3GPP Tables for link capacity derivation]
\label{sec:appendixA}
\begin{table}[ht]
\caption{Table 5.3.2-1 of TS 38.101-2 v17.6: Maximum transmission bandwidth configuration NRB : FR2}
 \label{tab:RE}
\centering
\begin{tabularx}{\columnwidth}{|X|X|X|X|X|X|X|X|}
\hline
 \textbf{$\mu$} & \textbf{50 MHz} & \textbf{100 MHz} & \textbf{200 MHz} & \textbf{400 MHz} & \textbf{800 MHz} & \textbf{1600 MHz} & \textbf{2000 MHz} \\ [1ex] \cline{2-8} 
 & $N_{\text{RB}}$ & $N_{\text{RB}}$ & $N_{\text{RB}}$ & $N_{\text{RB}}$ & $N_{\text{RB}}$ & $N_{\text{RB}}$ & $N_{\text{RB}}$ \\ \hline
 2&  66&  132&  264&  N/A&  N/A&  N/A&  N/A\\ \hline
 3&  32&  66&  132&  264&  N/A&  N/A&  N/A\\ \hline
 5&  N/A&  N/A&  N/A&  66&  124&  248&  N/A\\ \hline
 6&  N/A&  N/A&  N/A&  33&  62&  124&  148\\ \hline
\end{tabularx}
\end{table}
\begin{table}[ht]
\caption{Table 5.1.3.1-2 of 3GPP TS 38.214: Extended MCS index table 2 for PDSCH}
 \label{tab:MCS}
\centering
 \begin{tabular}{|c|c|c|c|c|} 
 \hline
 \textbf{SNR [dB]} & $\bm{I_{mcs}}$ & $\bm{Q_m}$ & \textbf{R} & \textbf{SE}\\ [1ex] 
 \hline
 $-1$ & 0 & 2 & 120 & 0.2344\\[1ex]
  0 & 1 & 2 & 193 & 0.377\\[1ex]
  1 & 2 & 2 & 308 & 0.6016\\[1ex]
  3 & 3 & 2 & 449 & 0.877\\[1ex]
  5 & 4 & 2 & 602 & 1.1758\\[1ex]
  7 & 5 & 4 & 378 & 1.4766\\[1ex]
  8 & 6 & 4 & 434 & 1.6953\\[1ex]
  9 & 7 & 4 & 490 & 1.9141\\[1ex]
  10 & 8 & 4 & 553 & 2.1602\\[1ex]
  11 & 9 & 4 & 616 & 2.4063\\[1ex]
  12 & 10 & 4 & 658 & 2.5703\\[1ex]
  13 & 11 & 6  & 466 & 2.7305\\[1ex]
  14 & 12 & 6  & 517 & 3.0293\\[1ex]
  15 & 13 & 6 & 567 & 3.3223\\[1ex]
  16 & 14 & 6 & 616 & 3.6094\\[1ex]
  17 & 15 & 6 & 666 & 3.9023\\[1ex]
  18 & 16 & 6 & 719 & 4.2129\\[1ex]
  19 & 17 & 6 & 772 & 4.5234\\[1ex]
  20 & 18 & 6 & 822 & 4.8164\\[1ex]
  21 & 19 & 6 & 873 & 5.1152\\[1ex]
  22 & 20 & 8 & 682.5 &5.332\\[1ex]
  23 & 21 & 8 & 711 & 5.5547\\[1ex]
  24 & 22 & 8 & 754 & 5.8906\\[1ex]
  25 & 23 & 8 & 797 & 6.2266\\[1ex]
  26 & 24 & 8 & 841 & 6.5703\\[1ex]
  27 & 25 & 8 & 885 & 6.9141\\[1ex]
  28 & 26 & 8 & 916.5 & 7.1602\\[1ex]
  29 & 27 & 8 & 948 & 7.4063\\[1ex]
  \hline
\end{tabular}
\end{table}}
\end{document}